\PassOptionsToPackage{table}{xcolor}
\documentclass[journal]{IEEEtran}
\IEEEoverridecommandlockouts
\usepackage{times,amsmath,color,amssymb,graphicx,epsfig,cite,psfrag,subfigure,balance}
\usepackage{svg}
\usepackage{caption}

\usepackage{capt-of}
\usepackage{amsfonts,pifont,enumerate,cases}
\usepackage{mathrsfs} 
\usepackage{xcolor} 
\usepackage{verbatim} 
\usepackage{bm}
\usepackage{cuted,stfloats}
\usepackage{algorithm}
\usepackage{algorithmic}

\usepackage{balance}
\usepackage{longtable}
\usepackage{blindtext}
\usepackage{multirow}
\usepackage{float}
\usepackage{threeparttable}
\usepackage{makecell}
\usepackage[utf8]{inputenc}
\usepackage{url}
\usepackage{booktabs}
\usepackage{amssymb}
\usepackage{bbding}
\usepackage{pifont}
\usepackage{wasysym}
\usepackage{utfsym}
\usepackage[algo2e,ruled,linesnumbered,lined,boxed,commentsnumbered]{algorithm2e}
\usepackage[
    colorlinks=true,
    linkcolor=green,
    citecolor=green,
    urlcolor=magenta
]{hyperref}

\newtheorem{lemma}{\upshape{\textbf{Lemma}}}
\newtheorem{assumption}{\upshape{\textbf{Assumption}}}
\newtheorem{proposition}{\upshape{\textbf{Proposition}}}

\begin{document}
\title{Antenna Positioning and Beamforming Optimization in MA Enabled Secure ISAC Systems: A Gradient-Based Meta Learning Approach}
\author{Zhendong Li, Yujie Zhao, Zhou Su, Xiao Tang, Zhiqing Wei, Ying Wang, and Wen Chen \thanks{Zhendong Li and Xiao Tang are with the School of Information and Communication Engineering, Xi'an Jiaotong University, Xi'an 710049, China (email: lizhendong@xjtu.edu.cn; tangxiao@xjtu.edu.cn). Yujie Zhao is with the School of Software Engineering, Xi’an Jiaotong University, Xi'an 710049, China (email: 2224111483@stu.xjtu.edu.cn). Zhou Su is with the School of Cyber Science and Engineering, Xi'an Jiaotong University, Xi'an 710049, China (email: zhousu@ieee.org). Zhiqing Wei and Ying Wang are with the State Key Laboratory of Networking and Switching Technology, Beijing University of Posts and Telecommunications, Beijing 100876, China (e-mail: weizhiqing@bupt.edu.cn; wangying@bupt.edu.cn). Wen Chen is with the School of Integrated Circuits, Shanghai Jiao Tong University, Shanghai 200240, China (e-mail: wenchen@sjtu.edu.cn).}\thanks{(Corresponding author: Zhou Su)}
\vspace{-1.5em}}
\maketitle
\thispagestyle{empty}


		
\begin{abstract}
    Integrated sensing and communications (ISAC) significantly improves spectral efficiency but introduces security risks regarding the interception of embedded communication signals. This paper proposes an movable antenna (MA)-enabled secure ISAC system that utilizes the spatial degrees of freedom of MA to mitigate these risks. Then, a problem is formulated to maximize the system secrecy rate by jointly optimizing antenna positioning, transmit beamforming, and artificial noise. However, the principal challenge arises from the non-convexity of the optimization problem and the strong coupling of the optimization variables. Generally, traditional optimization methods for this problem suffer from complex mathematical derivations, while existing deep learning approaches rely heavily on the training data distribution. To address these issues, we introduce a gradient-based meta learning (GML) algorithm, which works without pre-training and demonstrates favorable performance. Specifically, the algorithm establishes a neural network for each optimization variable, where the gradient of the objective function with respect to the variable serves as the input, and the output of the network determines the variable's update step. By handling the constraints and constructing penalty terms, the global loss function is used to guide the optimization process. Extensive numerical simulations confirm that the proposed algorithm achieves satisfactory performance in terms of both communication security and sensing capabilities.
\end{abstract}
		

\section{Introduction}
    \IEEEPARstart{A}{s} the development towards the sixth-generation (6G) wireless networks accelerates, the demand for ubiquitous connectivity and high-precision environmental perception is growing exponentially. To accommodate these  requirements within limited radio resources, integrated sensing and communication (ISAC) has emerged as a transformative paradigm\cite{introduction_ISAC1}. Unlike traditional systems, ISAC enables the co-design of signal waveforms and the sharing of  hardware platforms, thereby significantly enhancing both spectral and energy efficiency \cite{introduction_ISAC3}.  Consequently, ISAC is widely identified as a core technique for supporting emerging applications such as autonomous driving, smart cities, and extended reality\cite{introduction_ISAC2,introduction_ISAC4}. However, the communication data embedded in the radar sensing signals can be easily intercepted by potential eavesdropper \cite{ISAC2}. Therefore, secure ISAC systems deserve more attention.
        
     Currently, extensive researches have been devoted to explore secure ISAC systems\cite{SECUREISAC1,SECUREISAC4,SECUREISAC8,SECUREISAC11,SECUREISAC9,SECUREISAC12,SECUREISAC10,SECUREISAC14}. 
    The authors in \cite{SECUREISAC4} addressed the scenario where the sensing target acts as an eavesdropper by designing artificial noise beamforming to maximize the sum secrecy rate under sensing constraints. In \cite{SECUREISAC11}, authors proposed a sensing-assisted physical layer security framework that estimates eavesdroppers' directions to inform a joint optimization of secrecy rate and estimation accuracy. \cite{SECUREISAC12} investigated an ISAC-assisted secure mobile edge computing system where radar signals are exploited to simultaneously jam and sense eavesdroppers for user energy minimization.
    Meanwhile, a cooperative multi-BS framework was proposed to maximize sensing gain under security constraints via coordinated beamforming in \cite{SECUREISAC14}.
    It is worth noting that the aforementioned works are predominantly based on fixed position antennas (FPA). Since the element pattern of FPA depends on the fixed geometry of antenna elements and can only be manipulated via amplitude and phase weighting, FPA-based secure ISAC systems suffer from limited beamforming flexibility, confining their high-performance operation to specific spatial areas. Although strategies such as artificial noise injection and secure beamforming have been widely adopted to degrade the eavesdropper's channel quality, their effectiveness is fundamentally limited by the fixed spatial degrees of freedom (DoFs) of FPA. This hinders the system from realizing the full potential of secure ISAC performance.

    Recently, the movable antenna (MA), also known as the fluid antenna system (FAS), has been proposed as a novel antenna paradigm\cite{MA1}, which can address the limitations of FPA to a certain extent\cite{FAS1}. Unlike FPA, MA allow dynamic adjustment of antenna positions, which enables the system to actively shape the wireless propagation environment. To explore the potential of MA in enhancing wireless communication performance, extensive researches have been conducted from different perspectives, e.g., system model\cite{MA2,MA3}, antenna positioning\cite{MAPosition,MA_jointopt1,MA_jointopt2,MA_jointopt3} and hardware implementation\cite{MAhardware1,MAhardware2}. In \cite{MA2}, authors provided performance investigation and channel modeling for wireless networks utilizing MA. 
    Additionally, research attention has also been directed towards optimization issues in MA systems, especially towards antenna positioning, e.g., \cite{MAPosition,MA_jointopt1,MA_jointopt2,MA_jointopt3}. In particular, \cite{MAPosition} developed an antenna positioning algorithm based on graph theory to maximize the received signal power at the receiver. In contrast, \cite{MA_jointopt1,MA_jointopt2,MA_jointopt3} constructed an alternating optimization (AO) architecture, derived from convex optimization theory, for the joint optimization of antenna positioning and beamforming in MA systems. Furthermore, \cite{MAhardware1,MAhardware2} investigated the hardware implementation of MA systems. The capability of MA not only boosts communication performance but also improves the accuracy and resolution of sensing\cite{MAISAC1}. In \cite{MAlow1,MAlow2}, the authors primarily discussed the empowerment of the low-altitude domain by MA-ISAC systems. In addition, in \cite{MAISAC_new1,MAISAC_new2,FAS2,MAISAC5}, the authors focused on the joint optimization in MA-enabled ISAC systems. 
    Specifically, \cite{MAISAC_new1,MAISAC_new2} jointly optimized antenna positioning and beamforming using AO and successive convex approximation techniques to enhance sensing and communication performance. For the FAS, the joint optimization of antenna positioning and beamforming in FAS-enabled ISAC systems was investigated in \cite{FAS2}, and deep reinforcement learning was employed to address the highly coupled non-convex problem.
    Furthermore, considering specific transceiver architectures, \cite{MAISAC5} explored a full-duplex ISAC system with discrete antenna position constraints and proposed a binary particle swarm optimization scheme to minimize the total power consumption. In summary, benefiting from the higher spatial DoFs of MA, MA-enabled secure ISAC systems demonstrate superior performance.

    In addition, the introduction of MA has also brought some major challenges. Specifically, the antenna position variables are embedded within the highly non-linear phase terms of the field response vectors and are intricately coupled with the secure beamforming coefficients\cite{MAdif1}. This inherent non-linearity results in a challenging non-convex optimization problem \cite{MAISAC3,MAISAC2}, which is difficult to solve efficiently using conventional methods. Therefore, several data-driven learning algorithms have been employed to address these challenges, e.g., \cite{MADRL,Tang2025Deep,MADL}. Nevertheless, a common problem for data-driven learning paradigms is dependency on the training data distribution. Consequently, to maintain robust performance across scenarios with varying data distributions, adaptation techniques such as fine-tuning are typically required. Recently, a model-driven meta learning  has demonstrated its capability as a generalizable framework for solving non-convex optimization problems\cite{Meta2}. For instance, the authors in \cite{Meta1} utilized a meta learning approach to address the downlink beamforming optimization problem in multi-input single-output systems and compared with traditional weighted minimum mean square error algorithm\cite{WMMSE}. In reconfigurable intelligent surface-aided communications, \cite{Meta4} explored the use of meta learning for the joint optimization of precoding and phase-shifting matrices, integrating manifold concepts into the algorithmic framework. Given that research on MA-enabled secure ISAC is currently in its nascent stage and predominantly relies on conventional optimization methods, this paper introduces the gradient-based meta learning (GML) algorithm. Meanwhile, existing application scenarios of the GML framework in communication systems have not yet covered MA secure ISAC systems, and several existing solution strategies cannot support the direct application of the GML framework to such systems. Therefore, this paper focuses on developing processing strategies tailored to the specific constraints of MA secure ISAC systems, with the expectation that the GML framework can achieve superior performance in system optimization. By leveraging the model-driven characteristics and generalizability to non-convex problems of GML, we aim to address the optimization challenges encountered by other optimization algorithms.
        
	Motivated by the preceding discussion, this paper primarily considers a maximization of secrecy rate of MA-enabled secure ISAC systems. The dual-functional base station (BS) is equipped with MA, which is modeled as a continuously distributed linear array.  The objective is to maximize the secrecy rate through the joint optimization of the antenna positioning, transmit beamforming and artificial noise. Given that the problem is mathematically non-convex and the optimization variables are highly coupled, this paper proposes a GML optimization framework for the solution, building upon a series of constraint handling strategies. Since GML utilizes only local gradient information and the non-linear fitting capability of neural networks (NN) to determine the descent path, it does not rely on the convex geometry properties of the objective function. Meanwhile, as a model-driven algorithm, GML is not affected by the training data distribution. The key contributions of this paper are outlined as follows: 
    \begin{itemize}
        \item We construct the MA-enabled secure ISAC system, which leverages the higher spatial DoFs provided by MA to enhance system performance. In this paper, MA is modeled as a continuously distributed linear array. Then, an optimization problem to maximize the secrecy rate is formulated  via jointly optimizing the antenna positioning, transmit beamforming, as well as artificial noise for preventing information leakage. It is difficult to solve this problem directly, owing to the non-convex objective function and the high coupling among the optimization variables.
        \item To address the non-convex problem, this paper introduces the GML optimization framework tailored to the handling of specific constraints. More specifically, we first handle the constraints by mapping unconstrained variables to MA positions and artificial noise matrix via continuously differentiable functions, and by transforming sensing constraints into sub-problems. A global loss function is then derived by combining the remaining constraints with the objective function to guide the neural network training. Finally, under the proposed constraint-handling scheme, the GML algorithm optimizes antenna positioning, transmit beamforming, and artificial noise sequentially. 
        \item Furthermore, we provide numerical simulation results to evaluate the communication and sensing performance of the proposed algorithm. Regarding communication performance, the proposed algorithm is shown to significantly outperform the comparative baselines. In terms of sensing performance, we provide the joint sensing waveforms under various user distributions and depict the convergence of the sensing metric over epochs under a specific user distribution. The results demonstrate that, under various user distributions, the proposed algorithm consistently achieves the goal of maximizing the system secrecy rate while ensuring sensing quality.
        \end{itemize}
        
    The remainder of this paper is organized as follows: In Section \ref{sec2}, we present the MA-enabled secure ISAC system model and the formulation of secrecy rate maximization problem. Section \ref{sec3} presents the constraint handling strategies and the GML optimization algorithm framework, followed by an analysis of the algorithm's computational complexity and theory. In Section \ref{sec4}, numerical results are provided and discussed. Finally, Section \ref{sec5} concludes the paper.

    \textit{Notations:} $a$, $\mathbf{a}$ and $\mathbf{A}$ denote a scalar, a vector and a matrix, respectively. $\left({\cdot}\right)^\text{T}$ and $\left({\cdot}\right)^\text{H}$ denote transpose and conjugate transpose, respectively. 
    $\text{Rank}\left({\mathbf{A}}\right)$, 
    $\text{tr}\left({\mathbf{A}}\right)$,
    $\text{Null}\left({\mathbf{A}}\right)$ and $\text{Image}\left({\mathbf{A}}\right)$ denote rank, trace, kernel and image space of matrix $\mathbf{A}$, respectively. $||\mathbf{A}||_\text{F} $ denotes the Frobenius norm of $\mathbf{A}$. $\mathbf{A}\left({i,j}\right)$ represents the element of $\mathbf{A}$ in row $i$ and column $j$. 
    $\text{d}\left({\cdot}\right)$ and $\partial\left({\cdot}\right)$ denote the differential and the partial differential, respectively. $\mathbb{E}\left({\cdot}\right)$ denotes the expectation of a random variable. $\text{span}\left({\cdot}\right)$ denotes a linear space spanned by a set of vectors. 
    $\text{dim}\left({\cdot}\right)$ denotes the dimension of a linear space. $\mathcal{CN}$ denotes the complex Gaussian distribution. $\mathbb Z$, $\mathbb R$ and $\mathbb C$ represent the sets of integer, real and complex numbers, respectively. $\text{diag}\left({\mathbf{a}}\right)$ denotes a diagonal matrix where the entry at the $i$-th row and $i$-th column corresponds to the $i$-th element of vector $\mathbf{a}$. $\mathbf{I}_L$ is a identity matrix with size $L\times L$. 
		
	\section{System Model and Problem Formulation}\label{sec2}
\begin{figure}[t]
		\centering
        \includegraphics[width=1.0\linewidth]{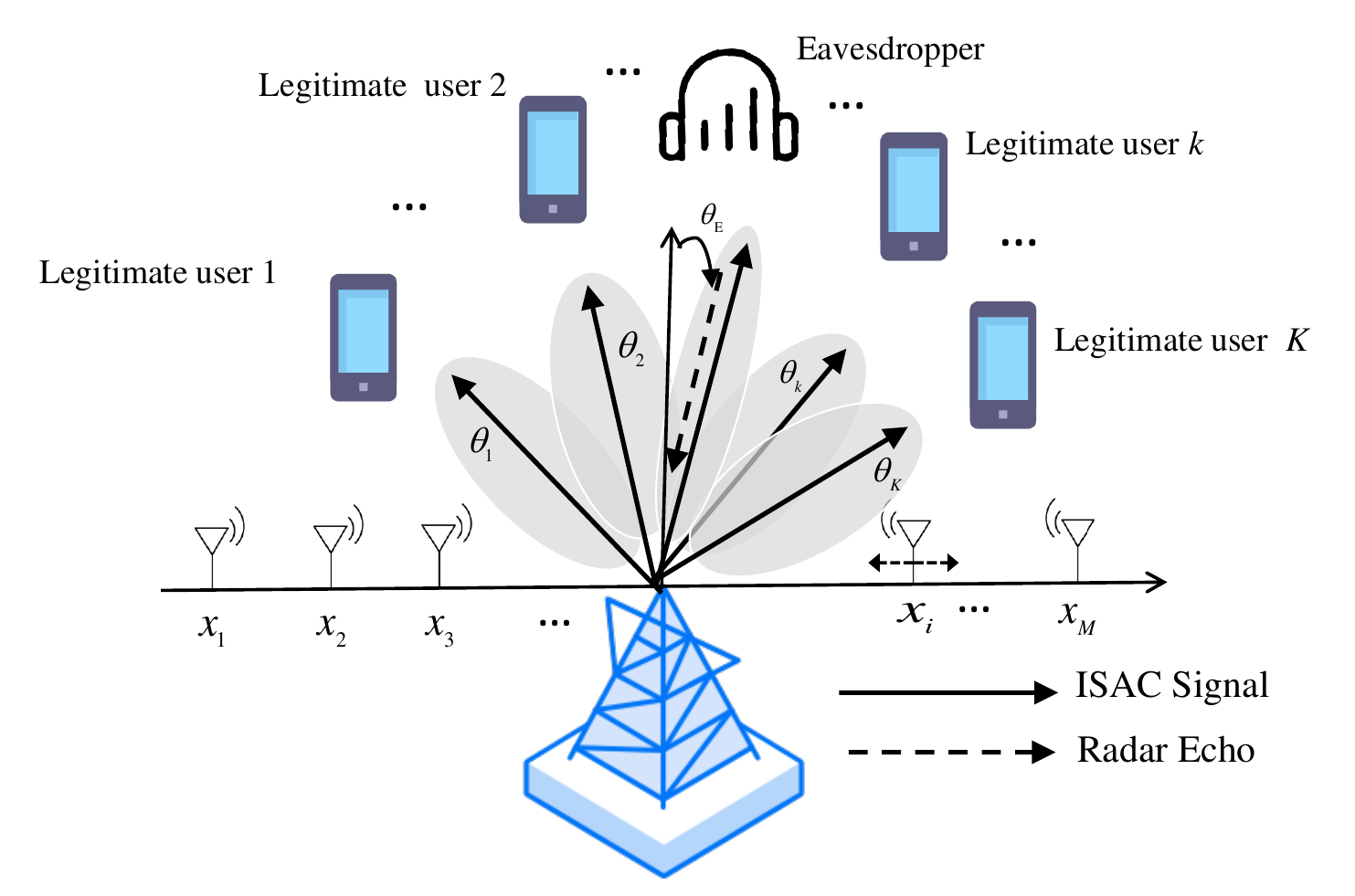}
			\caption{Illustration of MA enabled secure ISAC systems.}
			\label{fig:system model}
             \vspace{2em}
	\end{figure}

    In this paper, we consider a MA-enabled secure ISAC system, consisting of MA dual-functional radar and communication base station (BS), legitimate users and a target which is a potential eavesdropper. As illustrated in Fig.~\ref{fig:system model}, the BS is equipped with $M$ MA elements, capable of receiving echoes and estimating the user angles. Meanwhile, $K$ users are equipped with a single fixed antenna. The coordinates of the $m$-th transmit MA are denoted as \({{\bf{t}}_m} = {\left[ {{x_{m}},0} \right]^{\text{T}}}\), the coordinates of the $k$-th user are denoted as \({{\bf{r}}_k} = {\left[ {{x_{r,k}},{y_{r,k}}} \right]^{\text{T}}}\). The positions of all MA elements can be adjusted within a given one-dimensional area, which can be modeled as a straight line with length $\text{2}L$. Thus, the position constraint can be given by
    \begin{equation}
        \{x_m\}_{m=1}^M \in \left[-L, L\right].
    \end{equation}
    ${d_{\min }}$ is denoted as the minimum distance between any two MA elements, which means that any two position ${x_{s}}, {x_{c}}$ should satisfy the inequality below
    \begin{equation}
    \left| x_s - x_c \right| \ge d_{\text{min}}, \quad 1 \le s \ne c \le M.
    \end{equation}

	\subsection{Channel Model}
    For the considered system, the position set of $M$ antennas can be expressed as $\mathbf{p} = \left[ x_\text{1}, x_\text{2}, \ldots, x_M \right]^{\text{T}} \in \mathbb{R}^{M \times \text{1}}$. Therefore, given $\mathbf{p}$ and the steering angle with respect to the linear MA array as $\theta$, the corresponding steering vector of the MA array is given by
    \begin{equation}
        \boldsymbol{\alpha}(\mathbf{p}, \theta) =
            \left[
            e^{j\frac{2\pi}{\lambda}x_1 \sin(\theta)},
            \ldots,
            e^{j\frac{2\pi}{\lambda}x_M \sin(\theta)}
            \right]^{\text{T}} \in \mathbb{C}^{M \times 1}.
        \end{equation}
    The channels primarily include BS-user channels and BS-eavesdropper channel. For the BS-user channels, considering both line-of-sight (LoS) and non-line-of-sight (NLoS) factors between the BS and legitimate users, the channels are modeled as Rician fading channels. Denote the channel matrix as $\mathbf{H}_{\text{u}} = \left[ \mathbf{h}_\text{1}, \ldots, \mathbf{h}_K \right]^{\text{T}} \in \mathbb{C}^{K \times M}$. $\mathbf{h}_k \in \mathbb{C}^{M \times \text{1}}$ is denoted as the channel between BS and the $k$-th legitimate user, which can be given by
     \begin{equation}
        \mathbf{h}_k =
        \sqrt{C_0 \left( \frac{d(\mathbf{r}_k)}{D_0} \right)^{-o}} 
        \left(
        \sqrt{\frac{v_k}{1 + v_k}} \, \mathbf{h}_k^{\text{LoS}} +
        \sqrt{\frac{1}{1 + v_k}} \, \mathbf{h}_k^{\text{NLoS}}
        \right),
        \label{channel1}
    \end{equation}
    where ${C_\text{0}}$ is the path loss at the reference distance ${D_\text{0}}=\text{1}\text{m}$, $o$ represents the path loss exponent of the BS-user channel, $d(\mathbf{r}_k) = \left| \mathbf{r}_{\text{B}} - \mathbf{r}_k \right|$ represents  the distance between BS and the $k$-th legitimate user, ${v_k}$ represents the Rician factor of the channel. $\mathbf{h}_k^{\text{LoS}}$ can be equivalently represented by the steering vector $\boldsymbol{\alpha}(\mathbf{p}, \theta_k)$, where $\theta_k$ denotes the steering angle of the $k$-th legitimate user and ${\left[\mathbf{h}_k^{\text{NLoS}}\right]_m} \sim \mathcal{CN}(\text{0},\text{1})$. For BS-eavesdropper channel, considering that the direct path dominates the echo energy and carries the deterministic geometric information required for accurate parameter estimation, the channel  is modeled as LoS link, which can be given as
    \begin{equation}
        {\bf{h}}_\text{E}=\beta\bm\alpha({\mathbf{p}},\theta_\text{E}),
     \label{channel2}
    \end{equation}
where $\beta=\sqrt{C_\text{0}\left(\frac{d\left(\mathbf{r}_\text{E}\right)}{D_\text{0}}\right)^{-o}}$ represents the propagation loss.  
		\subsection{Signal Model}
        The signal vector ${\mathbf{x}} \in {\mathbb{C}^{M \times \text{1}}}$ transmitted from the BS to $K$ legitimate users can be given as
        \begin{equation}
            \mathbf{x} = \mathbf{W} \mathbf{s} + \mathbf{n},
        \end{equation}
        where $\mathbf{s} \in \mathbb{C}^{K \times \text{1}}$ denotes the information transmitted from the BS to  $K$ legitimate users and $\mathbb{E}[\mathbf{s}\mathbf{s}^{\text{H}} ]= \mathbf{I}$. $\mathbf{W} = \left[ \mathbf{w}_\text{1}, \ldots, \mathbf{w}_K \right] \in \mathbb{C}^{M \times K}$ denotes the beamforming matrix and $\mathbf{w}_k$ means the beamforming vector of the $k$-th legitimate user.  $\mathbf{n}$ denotes the artificial noise injected to prevent information leakage and $\mathbf{n} \sim \mathcal {CN} (\text{0},\mathbf{R}_{\text{N}})$, $\mathbf{R}_\text{N}=\mathbb E\left[\mathbf{n}\mathbf{n}^\text{H}\right]\in\mathbb{C}^{M\times{M}}.$ Therefore, for matrix $\mathbf{R}_\text{N}$, the following conditions should be satisfied: $\mathbf{R}_\text{N}=\mathbf{R}_\text{N}^\text{H}, \mathbf{R}_\text{N}\succeq \text{0}$.

        Then, the covariance matrix of the signal $\mathbf{x}$ can be given by
        \begin{equation}
            \mathbf{R}_{\text{X}} = \mathbb{E}\left[ \mathbf{x} \mathbf{x}^{\text{H}} \right] 
            = \sum_{k=1}^K \mathbf{w}_k \mathbf{w}_k^{\text{H}} + \mathbf{n} \mathbf{n}^{\text{H}}
            =\sum_{k=1}^K \mathbf{w}_k \mathbf{w}_k^{\text{H}} +\mathbf{R}_\text{N}.
        \end{equation}
        In addition, the transmit power is expressed as
        \begin{equation}
            P_{\text{t}} = \text{tr}\left( \mathbf{R}_{\text{X}} \right),
        \end{equation}
        the beampattern of steering angle $\theta$ is given by
        \begin{equation}
            P_{\text{bp}}=\bm\alpha^{\text{H}}(\mathbf{p},\theta)\mathbf{R}_{\text{X}}\bm\alpha(\mathbf{p},\theta).
        \end{equation}
        The signal vector received by the legitimate users $\mathbf{y} \in \mathbb{C}^{K\times \text{1}}$ can be given as
        \begin{equation}
            \mathbf{y} = \mathbf{H}_{\text{u}} \mathbf{x} + \mathbf{z},
        \end{equation}
        where $\mathbf{z}\in {\mathbb{C}^{K \times \text{1}}}$ is the noise vector of receivers and $\mathbf{z} \sim \mathcal{CN}\left(\text{0}, \sigma^\text{2} \mathbf{I}_K \right)$.

		\subsection{Performance Metrics}
        To evaluate the performance of the considered system, this paper define a number of performance metrics in this subsection included secure communication metrics and sensing metrics.
        \subsubsection{Secure Communication Metrics}
         Initially, based on the aforementioned system model, the signal-to-interference-plus-noise ratio (SINR) of the $k$-th legitimate user can be written as
         \begin{equation}
            \text{SINR}_k = \frac{
            \left| \mathbf{h}_k^{H} \mathbf{w}_k \right|^2
            }{
            \sum_{\substack{j=1 \\ j \neq k}}^{K} \left| \mathbf{h}_k^{H} \mathbf{w}_j \right|^2 
            + \mathbf{h}_k^\text{H}\mathbf{R}_\text{N}\mathbf{h}_k
            + \sigma^2
            }.
            \label{SINR}
        \end{equation}
        In the considered system model, the eavesdropper is assumed to possess perfect multiuser joint decoding or successive interference cancellation capabilities. Therefore, the SINR of the eavesdropper can be given by
         \begin{equation}
            \text{SINR}_{\text{E}} =
            \frac{
            |\beta|^2 \sum_{k=1}^{K} \left| {\bm \alpha}^{\text{H}}(\mathbf{p}, \theta_{\text{E}}) \mathbf{w}_k \right|^2
            }{
            |\beta|^2 \, \boldsymbol{\alpha}^{\text{H}}(\mathbf{p}, \theta_{\text{E}}) 
            \mathbf{R}_\text{N}
            \boldsymbol{\alpha}(\mathbf{p}, \theta_{\text{E}}) 
            + \sigma^2
            }.
        \label{SINRE}
        \end{equation}
        Thus, the system secrecy rate is defined as 
        \begin{equation}
            R_{\text{S}}= \left[\sum _{k=1}^K \log_2(1 + \text{SINR}_k)-\log_2(1+\text{SINR}_{\text{E}})
            \right]^{+}.
            \label{Rs}
        \end{equation}

        \subsubsection{Sensing Metrics}
        To achieve accurate and reliable sensing, it is necessary to employ energy focusing on the target region with low sidelobe levels, thereby effectively distinguishing the main signal reflections from environmental interference in echo analysis. Therefore, we discretize the angular domain $\left[ { - \frac{\pi }{\text{2}},\frac{\pi }{\text{2}}} \right]$ to $Q$ directions and define the ideal beam pattern $P_{\text{d}}\left(\theta_l\right)$ as
        \begin{equation}
            P_{\text{d}}(\theta_l) =
            \begin{cases}
            1, & \theta_t - \Delta \le \theta_l \le \theta_t + \Delta, \\
            0, & \text{others},
            \end{cases}
        \end{equation}
        where $\theta_t$ is the angel of the target and $\text{2}\Delta$ is the beamwidths for each target. Consequently, the mean squared error (MSE) between the ideal beampattern and the actual transmit beampattern is adopted to quantify the transmit-side sensing beampattern matching quality, which is given by
\begin{equation}
    \text{MSE}_{\text{bp}}=
    \frac{1}{Q} \sum_{q=1}^Q
    \left| \eta P_{\text{d}}(\theta_q)
    - \boldsymbol{\alpha}^{\text{H}}(\mathbf{p}, \theta_q)
    \mathbf{R}_{\text{X}}
    \boldsymbol{\alpha}(\mathbf{p}, \theta_q) \right|^2,
    \label{MSE_bp}
\end{equation}
where $\theta_q$ refers to the $q$-th sample point in the angular domain. 
The variable $\eta\geq \text{0}$ is an auxiliary scaling factor introduced to calibrate the amplitude level of the ideal sensing beampattern. 
Since $P_\text d(\theta_q)$ only specifies the desired angular 
support of the mainlobe and sidelobe regions, directly fixing its amplitude may 
introduce an artificial mismatch caused purely by the absolute transmit-power scale. 
Therefore, $\eta$ allows the ideal template to match the achievable beampattern level under 
the transmit-power budget. 
 

		\subsection{Problem Formulation}
	   Our objective is to maximize the secrecy rate among $K$ legitimate users by jointly optimizing antenna positioning, transmit beamforming, and artificial noise. This optimization is performed subject to constraints on the BS power budget, MA movement region and sensing metrics. Thus, the optimization problem can be formulated as 
		\begin{subequations}\label{optimization problem P0}
\renewcommand{\theequation}{16\alph{equation}} 
\begin{align}
    \text{P}_0: \max_{ \mathbf{p}, \mathbf{W}, \mathbf{R}_\text{N}, \eta}~ & R_{\text{S}}, \notag \\
    \text{s.t.} \quad
    & \text{tr}\left( \mathbf{R}_{\text{X}} \right) \le P_{\text{T}}, \label{P1 constraintt 1} \\
    &\mathbf{R}_\text{N}=\mathbf{R}_\text{N}^\text{H}, \mathbf{R}_\text{N}\succeq 0, \label{P1 constraintt 2}\\
    & \{x_{m}\}_{m=1}^M \in [-L,L], \label{P1 constraintt 3} \\
    & |x_{s} - x_{c}| \ge d_{\min},\quad 1 \le s \ne c \le M, \label{P1 constraintt 4} \\
    & \text{MSE}_{\text{bp}} \le \varepsilon, \label{P1 constraintt 5}
\end{align}
\end{subequations}
		where $P_{\text{T}}$ denots the maximun transmission power of the BS, and $\varepsilon$ is the pre-defined sensing metrics threshold. Specifically, constraint \eqref{P1 constraintt 1} specifies the BS transmit power limit, and constraint \eqref{P1 constraintt 2} guarantees the positive semi-definiteness of matrix $\mathbf{R}_\text{N}$. Constraints \eqref{P1 constraintt 3} \eqref{P1 constraintt 4} restrict the MA positions to the given panel region and ensure a minimum distance of $d_{\min}$ between any pair of MA elements, respectively. Constraint \eqref{P1 constraintt 5} serves as the sensing quality constraint, ensuring the system's sensing capability towards the eavesdropper. Due to the non-convexity of  objective function \eqref{Rs} and  constraints \eqref{P1 constraintt 4} \eqref{P1 constraintt 5}, as well as the coupling among multiple variables, this optimization problem is non-convex, posing significant challenges in obtaining optimal solutions. Thus, we employ the GML algorithm to solve this problem in the following sections.

		\section{Gradient-Based \\Meta Learning Optimization Algorithm}\label{sec3}
		
		In this section, considering the non-concavity of the objective function and the high coupling of the variables $\{\mathbf{p}, \mathbf{W}, \mathbf{R}_\text{N}, \eta\}$, we propose GML optimization algorithm to provide an adaptive updating strategy of each variable in problem \eqref{optimization problem P0}. The specific details of the constraint handling and GML execution procedure, are presented in the following subsections.
        By handling the constraints and incorporating penalty terms into the global loss function, the algorithm can find a high-quality solution within the feasible region.
       
		\subsection{Constraint Handling}
        GML can be viewed as a framework with a fixed execution pattern. The essential challenge in adapting this algorithm for diverse optimization problems is how to effectively integrate specific constraints into the framework using various processing methods. This subsection introduces the constraint handling strategies adopted herein, focusing on three primary constraint categories: MA position, sensing quality, and BS power budget. Adding a penalty term to the loss function is the most straightforward handling method. However, the penalty terms in the loss function influence the parameter update direction of the neural network by being combined with the gradient of the objective function. Hence, distinct handling strategies are employed in this paper for varying constraints.
        \subsubsection{MA Position Constraint}
        Given the geometric distribution of the MA array modeling in this paper, a variable substitution strategy is proposed to replace the constrained variable $\bf{p}$ with unconstrained variables $\bf{u}$. Specifically, in a linear MA array, optimizing the coordinates of the MA on the panel is equivalent to optimizing the spacing between the MA elements. For a panel of fixed dimensions, optimizing the spacing between MA elements is equivalent to determining a set of spatial allocation ratios for the panel. Considering the $M$ MA elements and the two endpoints of the panel, the number of spatial allocation ratios is $M+\text{1}$. Hence, this paper proposes to first obtain a set of panel spatial allocation ratios from the unconstrained variables ${\bf{u}}=\left[ {{u_\text{1}},...,{u_{M + \text{1}}}} \right]\in \mathbb R^{M+\text{1}}$ via a set of continuously differentiable mapping functions, which are then used to determine the MA position coordinates.

        Firstly, the vector of spatial allocation ratios is denoted as ${\bm{\pi}} = \left[ {{\pi_\text{1}},...,{\pi_{M + \text{1}}}} \right]\in \mathbb R^{M+\text{1}}$, where the elements of $\bm \pi$ must satisfy $\sum_{i=\text{1}}^{M+\text{1}}\pi_i=\text{1}$ and $\text{0} \le \pi_i \le \text{1},i=\text{1},\dots,M+\text{1}$. Therefore, the mapping function between $\bm{\pi}$ and $\bf{u}$ is constructed as follows
        \begin{equation}
            \bm{\pi}=\text{softmax}(\bf{u}),
            \label{position trf2}
        \end{equation}
        where ${\text{softmax}}\left(  \cdot  \right)$ is one of the most widely used functions in machine learning. It transforms a vector of $M+\text{1}$ arbitrary real values into a $M+\text{1}$ dimensional probability distribution. Therefore, the output of ${\text{softmax}}\left(  \cdot  \right)$ will strictly satisfy the constraints of $\bm{\pi}$. Specifically, $u_i$ is mapped through ${\text{softmax}}\left(  \cdot  \right)$ to obtain the corresponding $\pi_i$ as follows:
          \begin{equation}
            \text{softmax}(u_i)=\frac{{e^{u_i}}}{{\sum_{j=1}^{M+\text{1}}e^{u_j}}}=\pi_i.
            \label{mapping1}
        \end{equation}
        Additionally, ${\text{softmax}}\left(  \cdot  \right)$ is differentiable which means the gradient propagation will not be interrupted.
        
        Then, given \eqref{P1 constraintt 4}, the available spatial resources of the panel can be expressed as
        \begin{equation}
            L_{\text{a}}=2L-(M-1)d_{\text{min}}.
            \label{position trf1}
        \end{equation}
        The vector of space gap is denoted as ${\bf{g}}=[g_\text{1},...,g_{M+\text{1}}]\in \mathbb R^{M+\text{1}}$, which can be given by
        \begin{equation}
            {\bf{g}}=L_{\text {a}}\bm{\pi}
            \label{mapping2}.
        \end{equation}

        Lastly, the mapping function between $\bf{p}$ and $\bf{g}$ can be given as: $x_\text{1}=-L+g_\text{1}$, $x_\text{2}=x_\text{1}+g_\text{2}+d_\text{min}=(-L+g_\text{1})+g_\text{2}+d_\text{min}$..., without loss of generality, this mapping relationship can be expressed as
        \begin{equation}
        {x_m} = \left( { - L + {g_1}} \right) + \sum\limits_{k = 2}^{m} {{d_{{\rm{min}}}} + {g_k}} , m = 1, 2, ..., M,
          \label{mapping3}
        \end{equation}
        this mapping ensures that the resulting $\mathbf{p}$ necessarily satisfies the constraints \eqref{P1 constraintt 3} \eqref{P1 constraintt 4}. It is worth noting that, since the softmax mapping yields strictly positive allocation ratios, the resulting parameterization corresponds to the interior of the feasible set induced by the gap variables. Therefore, boundary configurations are not attained exactly, but they can be approached arbitrarily closely by choosing sufficiently large values in $\mathbf{u}$.

        Following the discussion above, a continuously differentiable mapping between the unconstrained variable $\bf{u}$ and the constrained variable $\bf{p}$ is established. In what follows, this paper will elaborate on the equivalence of this variable substitution in solving the optimization problem. To proceed, this paper will prove the following proposition
        \begin{equation}
        \begin{aligned}
        \text{P}_a: & \min_{\mathbf{p}} Z, \\
        \text{s.t. } & \mathbf{p} \in \mathbf{C},
        \end{aligned}
        \quad \Leftrightarrow \quad
        \begin{aligned}
        \text{P}_b: & \min_{\mathbf{u}} Z, \\
                    & f(\mathbf{u}) = \mathbf{p},
        \label{provement}
        \end{aligned}
        \end{equation}
        where $\text{P}_a$ and $\text{P}_b$ refers to 
        two simplified optimization problem, 
        $Z$ refers to the objective function, 
        $\mathbf{C}$ refers to the feasible region 
        of $\mathbf{p}$ and ${f}\left(  \cdot  \right)$ 
        refers to the mapping function between 
        $\mathbf{u}$ and $\mathbf{p}$ disucced above. 
        Here, the term $\Leftrightarrow$ 
        is used in the optimization sense for interior feasible points, i.e., the transformed problem preserves the feasible descent directions and stationary point characterization within the interior parameterization. For boundary points, although they are not represented exactly, their objective values can be approximated arbitrarily closely by interior points generated through the softmax transformation. The core of proving \eqref{provement} lies in the analysis of the Jacobian matrix ${\bf{J}}_f$. According to the mapping function between $\mathbf{u}$ and $\mathbf{p}$, $\mathbf{J}_f$ can be given as 
    \begin{equation}
    {{\mathbf{J}}_f} = \frac{{\partial {\mathbf{p}}}}{{\partial {\mathbf{u}}}} = \frac{{\partial {\mathbf{p}}}}{{\partial {\mathbf{g}}}} \frac{{\partial {\mathbf{g}}}}{{\partial {\bm{\pi }}}} \frac{{\partial {\bm{\pi }}}}{{\partial {\mathbf{u}}}}.
    \label{Jf}
    \end{equation}

   \begin{lemma}
    \label{Lemma1} 
     \textit{$\mathbf{J}_f$ has full row rank, which means for} $\mathbf{J}\in \mathbb R^{M\times (M+\text{1})}$, \textit{it holds that} $\text{Rank}(\mathbf{J}_f)=M$.
\end{lemma}

     \textit{Proof:} Please refer to Appendix \ref{appendixA}.$\hfill\blacksquare$
     
     Based on the analysis of $\mathbf{J}_f$, the following two propositions can be derived
     \begin{proposition}
    \label{Lemma2} 
     \textit{A gradient descent step in the $\mathbf{u}$-space induces a movement direction in the $\mathbf{p}$-space that aligns with the direction of gradient descent in the $\mathbf{p}$-space.}
    \end{proposition}
    
    \textit{Proof:} Please refer to Appendix \ref{appendixB}.$\hfill\blacksquare$
    
    Crucially, the gradient descent direction in the $\mathbf{u}$-space maps consistently to that in the $\mathbf{p}$-space. This implies that optimizing $\mathbf{u}$ in $\text{P}_b$ effectively drives $\mathbf{p}$ along its gradient descent direction. Therefore, the variable updates derived from $\text{P}_b$ are equivalent to those directly obtained from $\text{P}_a$.
    \begin{proposition}
    \label{Lemma3} 
     \textit{If there exists ${{\mathbf{u}}^ * }$ such that ${\nabla _{\mathbf{u}}}Z\left( {{{\mathbf{u}}^ * }} \right) = \text{0}$, then  ${\nabla _{\mathbf{p}}}Z = \text{0}$ and the ${{\mathbf{p}}^ * }$ that makes ${\nabla _{\mathbf{p}}}Z\left( {{{\mathbf{p}}^ * }} \right) = \text{0}$ satisfies $f\left( {{{\mathbf{u}}^ * }} \right) = {{\mathbf{p}}^ * }$}.
    \end{proposition}
    
     \textit{Proof:} Please refer to Appendix \ref{appendixC}.$\hfill\blacksquare$
     
     Furthermore, it can been demonstrated that updating $\mathbf{u}$ according to $\text{P}_b$'s gradient information and mapping it to $\mathbf{p}$ yields a zero gradient in the $\mathbf{p}$-space, which is equivalent to directly updating $\mathbf{p}$ based on $\text{P}_a$ gradient information. Therefore, $\text{P}_b$ can be considered equivalent to $\text{P}_a$, the proposition \eqref{provement} is proved. 

        \subsubsection{Sensing Quality Constraint}
        Given the mathematical complexity of \eqref{P1 constraintt 5}, we consider transforming this constraint. By leveraging its physical interpretation, a subproblem $\text{P}_\text{1}$ with fixed $\mathbf{p}$ is constructed as follows \cite{Rd}
        \begin{subequations}\label{optimization problem Rd}
        \renewcommand{\theequation}{24\alph{equation}} 
        \begin{align}
        \text{P}_1: \min_{ \mathbf{R}_\text{d} , \eta} & \frac{1}{Q} \sum_{q=1}^Q \left| \eta P_{\text{d}}(\theta_q) - 
        \boldsymbol{\alpha}^{\text{H}}(\mathbf{p}, \theta_q) \mathbf{R}_{\text{d}} \boldsymbol{\alpha}(\mathbf{p}, \theta_q) \right|^2, \notag \\
        \text{s.t.} \quad
        & \text{tr}\left( \mathbf{R}_{\text{d}} \right) = P_{\text{T}}, \label{P2 constraintt 1} \\
        & {{\bf{R}}_{\rm{d}}}\succeq0,{{\bf{R}}_{\rm{d}}} = {\bf{R}}_{\rm{d}}^{\rm{H}},\label{P2 constraintt 2} \\
        & \eta  \ge 0, \label{P2 constraintt 3}
    \end{align}
    \end{subequations}
    where $\eta$ is the scaling factor of \eqref{optimization problem P0}, $\mathbf{R}_{\text{d}}$ is defined as the desired beamforming matrix which satisfies the sensing constraint under the given conditions. Since both the objective function and the constraints in subproblem \eqref{optimization problem Rd} are convex\cite{ISAC2}, the CVX toolbox can be applied to obtain a high-quality solution. As a result, the optimized solution $\mathbf{R}_{\text{X}}$ in the original problem \eqref{optimization problem P0} can be made to closely approximate $\mathbf{R}_{\text{d}}$, which means \eqref{P1 constraintt 4} can be transformed as follows
    \begin{equation}
        \left\| {\mathbf{R}_{\rm{X}} - \mathbf{R}_{\rm{d}}} \right\|_\text{F}^2 \le \xi.
    \end{equation}
    where $\xi$ is the pre-defined threshold. The threshold $\xi$ characterizes the maximum allowable average beampattern mismatch. A smaller $\xi$ imposes a stricter target energy focusing and sidelobe suppression requirement, but also reduces the feasible region available for secrecy rate maximization. Hence, $\xi$ can be selected according to the sensing resolution requirement.

    \subsubsection{BS Transmit Power Budget Constraint} First, consider the semi-positive definite constraint on $\mathbf{R}_\text{N}$, i.e., \eqref{P1 constraintt 2}. Similar to the approach for handling MA position constraint, we first construct an unconstrained matrix $\mathbf{N}\in\mathbb C^{M\times M}$, optimize $\mathbf{N}$, and then obtain $\mathbf{R}_\text{N}$ via ${{\mathbf{R}}_{\text{N}}} = {\mathbf{N}}{{\mathbf{N}}^{\text{H}}}$.
    Based on all the discussion above, \eqref{optimization problem P0} can be written as a simplified problem $\text{P}_\text{2}$ as follows
\begin{subequations}\label{optimization problem P2}
\renewcommand{\theequation}{26\alph{equation}} 
\begin{align}
    \text{P}_2: \max_{ \mathbf{p}, \mathbf{W}, \mathbf{N}}~ & R_{\text{S}}, \notag \\
    \text{s.t.} \quad
    & \text{tr}\left( \mathbf{R}_{\text{X}} \right) \le P_{\text{T}}, \label{P2 constraint 1} \\
    &\left\| {\mathbf{R}_{\rm{X}} - \mathbf{R}_{\rm{d}}} \right\|_\text{F}^2 \le \xi.\label{P2 constraint 2}
\end{align}
\end{subequations}
The penalty terms corresponding to \eqref{P2 constraint 1} \eqref{P2 constraint 2} are denoted as $\mathcal L_\text{1}$ and $\mathcal L_\text{2}$. Then, the global loss function can be constructed as 
\begin{equation}
    {\cal L} =  - R_\text{S} + {\lambda _1}{{\cal L}_1} + {\lambda _2}{{\cal L}_2},
\end{equation}
where ${\lambda _\text{1}}$ and ${\lambda _\text{2}}$ represent the weights of the penalty terms, respectively. $\mathcal L_\text{1}$ is defined as
\begin{equation}
    {\cal L}_1 = 
    \begin{cases}
        \text{tr}\left( \mathbf{R}_{\rm{X}} \right)-P_{\rm{T}} , & \text{if } \text{tr}\left( \mathbf{R}_{\rm{X}} \right) > P_{\rm{T}}, \\
        0, & \text{if } \text{tr}\left( \mathbf{R}_{\rm{X}} \right) \le P_{\rm{T}}.
    \end{cases}
\end{equation}
$\mathcal L_\text{2}$ is defined as
\begin{equation}
    {\cal L}_2 = 
    \begin{cases}
        \left\| \mathbf{R}_{\rm{X}} - \mathbf{R}_{\rm{d}} \right\|^2-\xi , & \text{if } \left\| \mathbf{R}_{\rm{X}} - \mathbf{R}_{\rm{d}} \right\|^2 > \xi, \\
        0, & \text{if } \left\| \mathbf{R}_{\rm{X}} - \mathbf{R}_{\rm{d}} \right\|^2 \le \xi.
    \end{cases}
\end{equation}
 By transforming the constraints and incorporating penalty terms into the global loss function, the algorithm can find a high-quality solution within the feasible region.

\subsection{GML Optimization Algorithm}
In this subsection, we will elaborate on this algorithm in detail from three aspects, included the gradient-based input mechanism, meta learning  framework and basic meta learning NNs. 
\subsubsection{Gradient Input Mechanism}
\begin{figure*}[t]
    \centering
\includegraphics[width=1.0\textwidth]{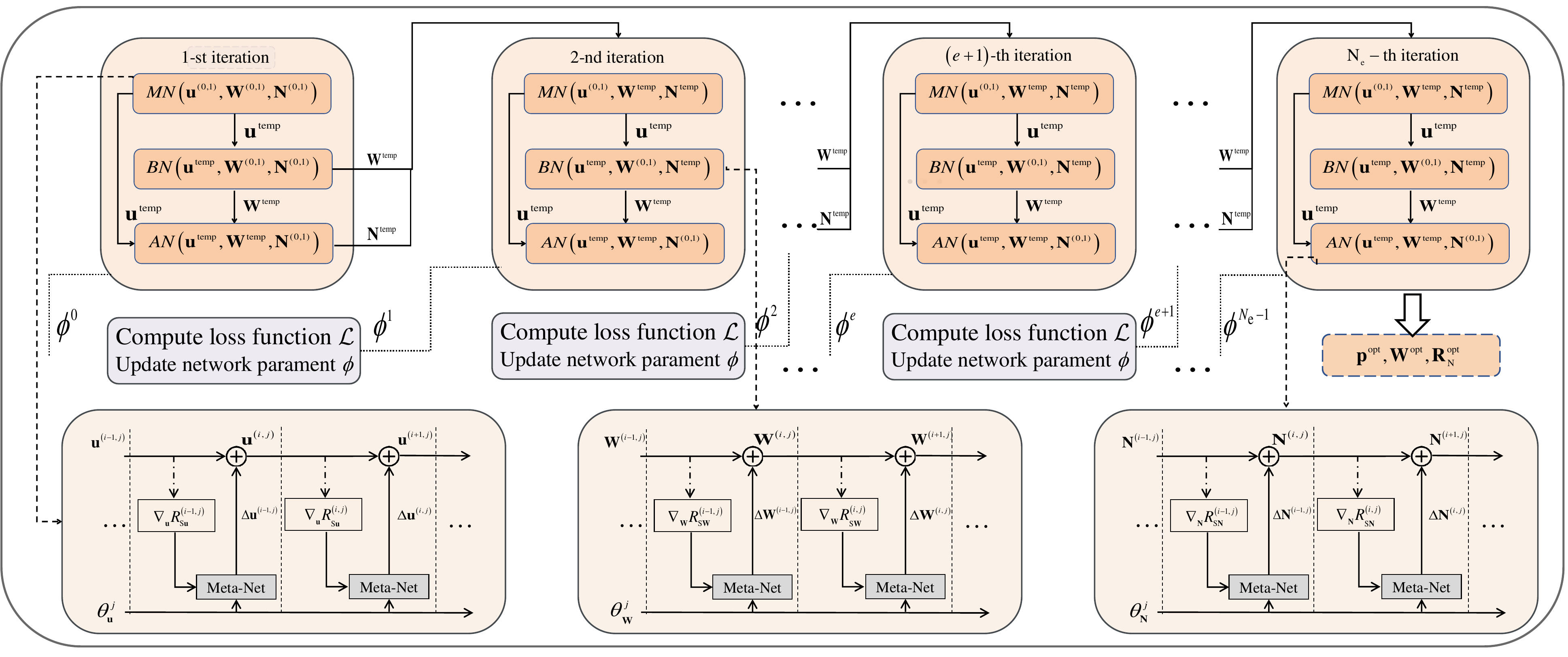}
    \caption{Overall framework of the GML optimization algorithm.}
    \label{fig:flowchart}
    \vspace{2em}
\end{figure*}

For the gradient-based input mechanism, as previously mentioned, this algorithm constructs a neural network for each variable. This network takes the gradient of the objective function with respect to the optimization variable as input, and its output serves as the update step size for that variable, which can be described in Fig.~\ref{fig:flowchart}. Compared to optimization modes that rely solely on variable inputs and variable outputs, the gradient-based input mechanism, on one hand, demonstrates superior optimization performance by leveraging the high-dimensional information within the gradient. On the other hand, it enhances the algorithm's interpretability, as it is no longer a mere black box with only inputs and outputs. To elaborate, in traditional gradient descent algorithms, the update step for an optimization variable is expressed as the learning rate multiplied by the gradient of the objective function with respect to that variable. In other words, the update step can be viewed as a function of the gradient. Concurrently, it has been proven that neural networks are, in theory, universal approximators capable of fitting any arbitrary function. Therefore, this gradient-based input mechanism can be interpreted as follows: the neural network corresponding to each variable is, in effect, fitting this update step function under the current data distribution through successive updates of its network parameters. This function takes the gradient as its input and yields the update step as its output.

\begin{algorithm}[t]
\caption{GML Optimization Framework }
\label{Algorithm 1}

Randomly initialize $\phi_{\mathbf{u}}, \phi_{\mathbf{W}}, \phi_{\mathbf{N}}, \mathbf{u}^{(\text{0},\text{1})}, \mathbf{W}^{(\text{0},\text{1})}, \mathbf{N}^{(\text{0},\text{1})}$;\\
Initialize the maximum recorder as $\text{MAX}=\text{0}$.

\For{$e = \text{1}$ \KwTo $N_e$}{
    $\bar{\mathcal{L}}=\text{0}$;
    
    \For{$j = \text{1}$ \KwTo $N_o$}{
        $\mathbf{u}^{(\text{0},j)} = \mathbf{u}^{(\text{0},\text{1})}$,\;
        $\mathbf{W}^{(\text{0},j)} = \mathbf{W}^{(\text{0},\text{1})}$,\;
        $\mathbf{N}^{(\text{0},j)} = \mathbf{N}^{(\text{0},\text{1})}$;\;
        
        \For{$i = \text{1}$ \KwTo $N_i$}
        {
            $R^{(i-\text{1},j)}_{\text{S}\mathbf{u}} = R_\text{S}(f(\mathbf{u}^{(i-\text{1},j)}), \mathbf{W}^{\text{temp}}, \mathbf{N}^{\text{temp}})$;\\
            $\Delta \mathbf{u}^{(i-\text{1},j)} = \text{NN}_{\mathbf{u}}(\nabla_{\mathbf{u}} R^{(i-\text{1},j)}_{\text{S}\mathbf{u}})$;\\
            $\mathbf{u}^{(i,j)} \leftarrow \mathbf{u}^{(i-\text{1},j)} + \Delta \mathbf{u}^{(i-\text{1},j)}$.
        }
        
        ${{\bf{u}}^{\text{temp}}} = \mathbf{u}^{(N_i,j)}$;\\
        ${{\bf{p}}^{\text{temp}}} = f\left( {\bf{u}}^{\text{temp}} \right)$;\\
        Update channel according to ${\bf{p}}^{\text{temp}}$ based on \eqref{channel1} \eqref{channel2};
        
        \If{$e \mid n_\text{0}$}{
            Set $\mathbf{R}_\text{d}$ according to problem~\eqref{optimization problem Rd}.\;
        }
        
        \For{$i = \text{1}$ \KwTo $N_i$}{
            $R^{(i-\text{1},j)}_{\text{S}\mathbf{W}} = R_\text{S}(\mathbf{p}^{\text{temp}}, \mathbf{W}^{(i-\text{1},j)}, \mathbf{N}^{\text{temp}})$;\\
            $\Delta \mathbf{W}^{(i-\text{1},j)} = \text{NN}_{\mathbf{W}}(\nabla_{\mathbf{W}} R^{(i-\text{1},j)}_{\text{S}\mathbf{W}})$;\\
            $\mathbf{W}^{(i,j)} \leftarrow \mathbf{W}^{(i-\text{1},j)} + \Delta \mathbf{W}^{(i-\text{1},j)}$.
        }
        
        $\mathbf{W}^{\text{temp}} = \mathbf{W}^{(N_i,j)}$;
        
        \For{$i = \text{1}$ \KwTo $N_i$}{
            $R^{(i-\text{1},j)}_{\text{S}\mathbf{N}} = R_\text{S}(\mathbf{p}^{\text{temp}}, \mathbf{W}^{\text{temp}}, \mathbf{N}^{(i-\text{1},j)})$;\\
            $\Delta \mathbf{N}^{(i-\text{1},j)} = \text{NN}_{\mathbf{N}}(\nabla_{\mathbf{N}} R^{(i-\text{1},j)}_{\text{S}\mathbf{N}})$;\\
            $\mathbf{N}^{(i,j)} \leftarrow \mathbf{N}^{(i-\text{1},j)} + \Delta \mathbf{N}^{(i-\text{1},j)}$.
        }
        
        $\mathbf{N}^{\text{temp}} = \mathbf{N}^{(N_i,j)}$;\\
        $\mathbf{R}_\text{N}^\text{temp}=\mathbf{N}^\text{temp}(\mathbf{N}^\text{temp})^\text{H}$;\\
        $R_\text{S}^j=R_\text{S}(\mathbf{p}^{\text{temp}},
        \mathbf{W}^{\text{temp}}, \mathbf{R}_\text{N}^{\text{temp}})$;\\
        $\mathcal{L}^j = -R_\text{S}^j+{\lambda _\text{1}}{{\cal L}_\text{1}^j} + {\lambda _\text{2}}{{\cal L}_\text{2}^j}$;\\
        $\bar{\mathcal{L}} = \bar{\mathcal{L}} + \mathcal{L}_j$;
        
        \If{$R_\text{S}^j > \text{MAX}$}{
            $\text{MAX} = R_\text{S}^j$;\\
            $\mathbf{p}^{\text{opt}}= \mathbf{p}^{\text{temp}}$,\;
            $\mathbf{W}^{\text{opt}}= \mathbf{W}^{\text{temp}}$,\;
            $\mathbf{R}_\text{N}^{\text{opt}}= \mathbf{R}_\text{N}^{\text{temp}}$.\;
        }
    } 
    
    $\bar{\cal L} = \frac{\text{1}}{N_o} \bar{\cal L}$;\\
    Update $\phi_\mathbf{p}$ as \eqref{update 1};
    
    \If {$e\mid n_\text{1}$}{
        Update $\phi_{\mathbf{W}}$ as \eqref{update 2};\\
        Update $\phi_\mathbf{N}$ as \eqref{update 3}.
    }
} 

\KwRet $\mathbf{p}^{\text{opt}}$, $\mathbf{W}^{\text{opt}}$, $\mathbf{R}_\text{N}^{\text{opt}}$.

\end{algorithm}
\subsubsection{GML Framework}
Benefiting from the superior learn to learn capability of the meta learning architecture, this learned update step function, however, is distinct from the simplest or adaptive-learning-rate gradient descent algorithms. This is because each update step function (which can be viewed as an optimizer for each variable) is iteratively optimized based on the data distribution under the current optimization conditions. Consequently, the optimizer corresponding to each optimization variable is different for different data distributions. In traditional CVX or machine learning (deep learning) based optimization algorithms, it can also be considered that an optimizer is constructed for each optimization variable. However, the parameters of these optimizers are often preset or learned based on the data distribution of the training set. During the execution of the optimization task, these parameters are generally considered fixed and immutable. In contrast, the GML optimization algorithm, as a model-driven meta learning algorithm, requires no presetting of optimization parameters, nor is its optimization performance dependent on the distribution of training data. By iterating the optimizer parameters based on the current optimization scenario, it exhibits superior generalization performance.

In summary, the GML optimization algorithm can be viewed as consisting of two parts: variable updates and optimizer updates. This section will elaborate on these two parts by decomposing them into three loops, which are denoted, from the innermost to the outermost, as the inner loop, the outer loop, and the epoch loop. The corresponding numbers of iterations are denoted as $N_\text{i}$, $N_\text{o}$ and $N_\text{e}$, respectively. The overall optimization framework can be summarized as {\bf{Algorithm}} \ref{Algorithm 1}.

The inner loop primarily performs cyclic updates of the optimization variables, as shown in Fig.~\ref{fig:flowchart}, there are three sub-networks responsible for optimizing $\mathbf{u}$, $\mathbf{W} $ and $\mathbf{N}$, including the \textit{Movable Antenna Position Network (MN)}, \textit{Beamforming Network (BN)}, and \textit{Artificial Noise Network (AN)}. In each optimization network, the variable to be optimized inherits its initialized value, while the other variables inherit the output results from their respective optimization networks. During the $j$-th outer loop and $i$-th inner loop, the variable update process can be expressed as
\begin{equation}
    {\mathbf{u}}^{\left( i+1, j \right)} = {\mathbf{u}}^{\left( i, j \right)} + \Delta {\mathbf{u}}^{\left( i, j \right)},
\end{equation}
\begin{equation}
    \mathbf{W}^{\left( i + 1, j \right)} = \mathbf{W}^{\left( i, j \right)} + \Delta \mathbf{W}^{\left( i, j \right)},
\end{equation}
\begin{equation}
    \mathbf{N}^{\left( i + 1, j \right)} = \mathbf{N}^{\left( i, j \right)} + \Delta \mathbf{N}^{\left( i, j \right)},
\end{equation}
where $i=\text{1},\cdots,N_\text{i},j=\text{1},\cdots,N_\text{o},$ ${\mathbf{u}}^{\left( i, j \right)}$, $ \mathbf{W}^{\left( i, j \right)}$ and $\mathbf{N}^{\left( i, j \right)}$  respectively represents the states of the optimization variables at the end of the $i$-th inner loop during the $j$-th outer loop. Specifically, ${\mathbf{u}}^{\left( \text{0}, j \right)}$, $ \mathbf{W}^{\left( \text{0}, j \right)}$ and $\mathbf{N}^{\left( \text{0}, j \right)}$ respectively represent the initial states of the variables. $\Delta {\mathbf{u}}^{\left( i, j \right)}$, $\Delta \mathbf{W}^{\left( i, j \right)}$ and $\Delta \mathbf{N}^{\left( i, j \right)}$ respectively represents the updated  step of the variables at the end of the $i$-th inner loop during the $j$-th outer loop. 
After $N_\text{i}$ inner epochs, the temporarily optimized variables is given by
\begin{equation}
    {{\bf{p}}^\text{temp}} = f\left({{\bf{u}}^\text{temp}}\right),
\end{equation}
\begin{equation}
    {\bf{W}}^\text{temp} = {\bf{W}}^{\left( N_i, j \right)},
\end{equation}
\begin{equation}
\mathbf{R}_\text{N}^\text{temp}=\mathbf{N}^\text{temp}(\mathbf{N}^\text{temp})^\text{H}.
\end{equation}

The outer loop is primarily responsible for computing the loss function based on the temporarily optimized variables, thereby implementing an unsupervised optimization strategy. The loss function in the $j$-th outer loop can be expressed as
\begin{equation}
    {{\cal L}^j} =  - R_\text{s}\left( {{{\bf{p}}^\text{temp}},{\bf{W}}^\text{temp},\mathbf{R}_\text{N}^\text{temp}} \right) + {\lambda _1}{\cal L}_1^j + {\lambda _2}{\cal L}_2^j.
\end{equation}

The epoch loop is responsible for updating the neural network parametes after the completion of the outer loop. After $N_\text{o}$ epochs, $\bar {\cal L}$ can be expressed as 
\begin{equation}
    \bar{\cal L} = \frac{1}{{N_o}} \sum_{j=1}^{N_o} {\cal L}^j.
\end{equation}
The parameter update direction of three sub-networks aims to minimize the $\bar {\cal L}$, which corresponds to maximizing the objective function within the constraint boundaries. The paraments of neural networks can be given as follows
\begin{equation}
    \phi_{\bf{u}}^{e + 1} = \phi_{\bf{u}}^{e} + \alpha_{\bf{u}} \, {\rm Adam} \left( \nabla_{\phi_{\bf{u}}^{e}} \bar{\cal L}, \phi_{\bf{u}}^{e} \right),
    \label{update 1}
\end{equation}
\begin{equation}
    \phi_{{\bf{W}}}^{e + 1} = \phi_{{\bf{W}}}^e + \alpha_{{\bf{W}}} \, {\rm Adam} \left( \nabla_{\phi_{{\bf{W}}}^e} \bar{\cal L}, \phi_{{\bf{W}}}^e \right),
     \label{update 2}
\end{equation}
\begin{equation}
    \phi_{\bf{N}}^{e + 1} = \phi_{\bf{N}}^e + \alpha_{\bf{N}} \, {\rm Adam}\left( \nabla_{\phi_{\bf{N}}^e} \bar{\cal L}, \phi_{\bf{N}}^e \right),
     \label{update 3}
\end{equation}
where $\phi_{\bf{u}}^{e}$, $\phi_{{\bf{W}}}^e$ and $ \phi_{\bf{N}}^e$ represent the parament set of each neural network during $e$-th epoch loop, $\alpha_{\bf{u}}$, $\alpha_{{\bf{W}}}$ and $\alpha_{\bf{N}}$ are the corresponding learning rate of neural networks. The parameters of each network are updated according to their respective update intervals, thereby balancing the alternating optimization process.

\subsubsection{Basic NNs of Meta Learning}
For \textit{MN}, as illustrated in \eqref{provement}, the algorithm just optimize $\mathbf{u}$ based on the equivalence between $\mathbf{u}$ and $\mathbf{p}$ in the optimization problem. Denote the secrecy rate in $i$-th inner loop and $j$-th outer loop as $R_{\text{S}\mathbf{u}}^{\left({i,j}\right)}$, which can be expressed as 
\begin{equation}
    R^{(i,j)}_{\text{S}\mathbf{u}} = R_\text{S}(f(\mathbf{u}^{(i,j)}), \mathbf{W}^{\text{temp}}, \mathbf{N}^{\text{temp}}),
\end{equation}
where $\mathbf{W}^{\text{temp}}$ and $\mathbf{N}^{\text{temp}}$ are either the initialized or updated. After one inner loop, $\mathbf{u}^\text{temp}$ is obtained, and the constraint-satisfying $\mathbf{p}^\text{temp}$ is derived via the mapping function. Subsequently, the channel is updated based on $\mathbf{p}^\text{temp}$ according to \eqref{channel1} \eqref{channel2} for the subsequent optimization of $\mathbf{W}$ and $\mathbf{N}$. The specific procedures in \textit{BN} and \textit{AN} are similar to that in \textit{MN}, and thus will not be elaborated upon here. The specific architectures of three networks are detailed in {\bf{Table}} \ref{tab:table1}. Considering the computational cost, the networks adopted in this paper are all simple multi-layer perceptrons, yet they still achieve superior performance.
\begin{table}[t] 
    \vspace{3.5em}
    \centering 
    \caption{Number of Neurons in the NNs}
    \label{tab:table1} 
    \begin{tabular}{clccc}
        \toprule
        No. & Layer Name & \textit{MN} & \textit{BN} & \textit{AN} \\
        \midrule
        1 & Input Layer & $M+\text{1}$ & 2$ \times K$ & 2$\times M$ \\
        2 & Linear Layer 1 & 100 & 200 & 100 \\
        3 & ReLU Layer & 100 & 200 & 100\\
        4 & Output Layer & $M+\text{1}$ & 2$ \times K$ & 2$\times M$ \\
        \bottomrule
    \end{tabular}
\end{table}

The proposed optimization framework in this paper ensures that, during each outer loop, each variable within the respective networks is optimized from scratch. The inner loops within each outer loop correspond to optimization steps, collectively forming an optimization trajectory. In this framework, the trajectory is continuously updated and refined, enabling the learning of an effective strategy.
\subsection{Complexity and Theoretical Analysis}
\subsubsection{Computational Complexity Analysis} The complexity of the proposed GML is analyzed considering the three-layer loop structure and the specific operations within each stage.
\begin{itemize}
\item {\bf{Inner Loop}}: In each of the $N_\text{i}$ inner iterations, the following steps are performed:

{\textit{(a) Calculation for Secrecy Rate}}: For a specific user $k$, according to \eqref{SINR}, calculating the desired signal power $| \mathbf{h}_k^H \mathbf{w}_k |^\text{2}$ involves a vector inner product of dimension $M$, with a complexity of $\mathcal{O}(M)$. Meanwhile, computing the interference power  $ \sum_{\substack{j \neq k}}^{K} \left| \mathbf{h}_k^{H} \mathbf{w}_j \right|^\text{2}$  requires inner products between user $k$'s channel and the other $K-\text{1}$ beamforming vectors. Therefore, the complexity for one user is $\mathcal{O}(KM)$, and repeating this for all $K$ users results in a total complexity of $\mathcal{O}(K^\text{2} M)$. Regarding the eavesdropper, according to \eqref{SINRE}, calculating $R_\text{E}$ involves evaluating the received power from all $K$ beamforming vectors, incurring a complexity of $\mathcal{O}(KM)$.

{\textit{(b) Mapping Function in \textit{MN}}}: In \textit{MN} process, the mapping function $f(\cdot)$ must be calculated before computing the secrecy rate.  Since \eqref{mapping1} \eqref{mapping2} \eqref{mapping3} these operations involve linear scanning of the $M$-dimensional vector, the complexity is $\mathcal{O}(M)$.

Owing to the automatic differentiation mechanism, the computational overhead for deriving gradients is proportional to the forward propagation. Thus, it does not increase the asymptotic complexity order. Since the embedded NNs are relatively
small and shallow, the computational complexity of the NNs
is approximately $\mathcal{O}(M)$, $\mathcal{O}(KM)$, $\mathcal{O}(K^\text{2})$for \textit{MN, BN, AN}. Denote all of them as $\mathcal{O}(\text{NNs})$. Thus, total computational complexity can be expressed as $\mathcal{O}(K^\text{2} M)+\mathcal{O}(KM)+\mathcal{O}(M)+\mathcal{O}(\text{NNs})=\mathcal{O}(K^\text{2} M)$.

\item {\bf{Outer Loop}}: In each of the $N_\text{o}$ outer iterations, the following steps are performed:

{\textit{(a) Channel Reconstruction}}: Since the channel state information depends on the antenna positions, the channel vectors for all $K$ legitimate users and the eavesdropper must be reconstructed after each position update. Calculating the steering vector for $M$ antennas for $K+\text{1}$ nodes incurs a complexity of $\mathcal{O}(KM)$.

{\textit{(b) Update of} }$\bf{R}_\text{d}$: 
Every $n_\text{0}$ epochs, the desired 
covariance matrix $\bf{R}_\text{d}$ is updated by solving the 
convex optimization problem \eqref{optimization problem Rd}. Solving 
this problem with $M \times M$ variables using standard interior-point methods 
typically incurs a complexity of $\mathcal{O}(M^{\text{3.5}})$. Over the entire 
training process, this update is performed $\frac{N_e}{n_\text{0}}$ times. Thus, 
the total complexity for this stage is $\mathcal{O}\left( \frac{N_e}{n_\text{0}} M^{\text{3.5}} \right)$.

{\textit{(c) Loss Function}}: The loss includes the secrecy rate (analyzed above as $\mathcal{O}(K^\text{2} M)$) power penalty terms ($\mathcal{O}(KM)$) and sensing penalty terms $\mathcal{O}(K M^\text{2})$. 

 Thus, total computational complexity of outer loop can be expressed as $\mathcal{O}(KM)+\mathcal{O}(K^\text{2}M)+\mathcal{O}(KM)+\mathcal{O}(KM^\text{2})=\mathcal{O}(KM^\text{2})$.

Considering the number of the inner, outer and epoch
iterations, the overall complexity of the proposed GML
algorithm is $\mathcal{O}\left( \frac{N_e}{n_\text{0}} M^{\text{3.5}} + N_e N_o \left( N_i K^\text{2} M + K M^\text{2} \right) \right)$.

\end{itemize}

\begin{figure}[t]
    \centering
    \includegraphics[width=0.85\columnwidth]{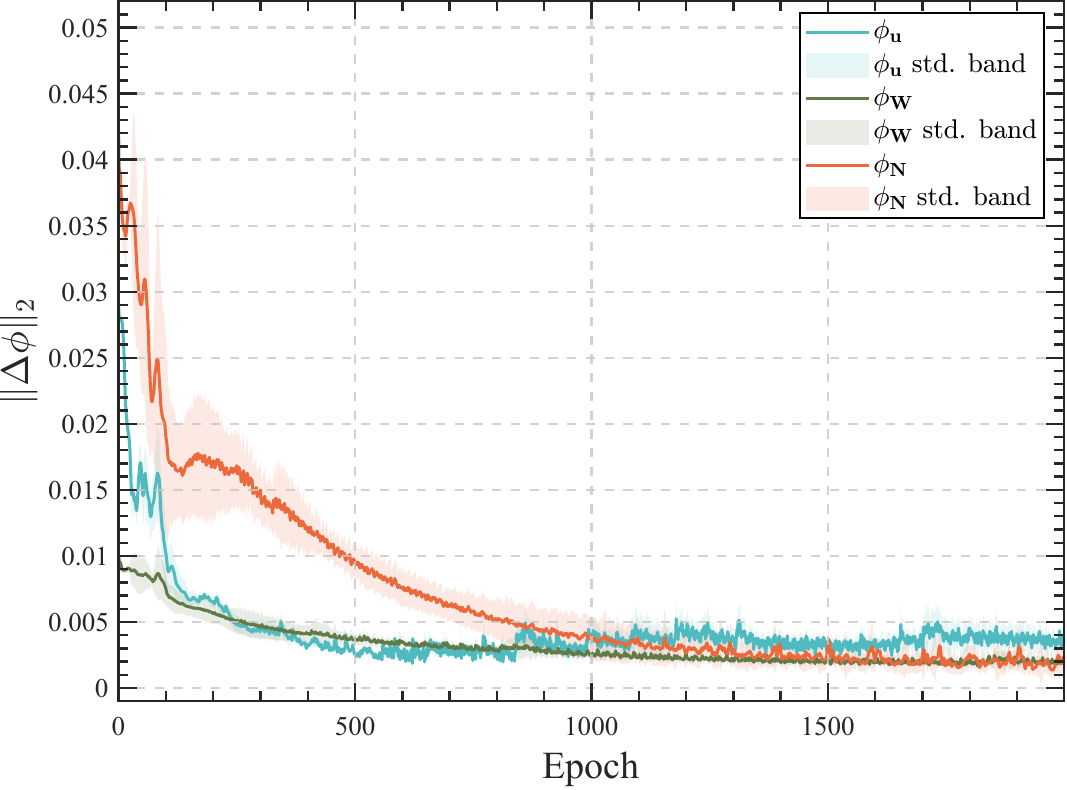}
    \caption{Evolution of the successive meta-parameter variations
    $\|\phi_u^{(e+\text{1})}-\phi_u^{(e)}\|_\text{2}$,
    $\|\phi_W^{(e+\text{1})}-\phi_W^{(e)}\|_\text{2}$, and
    $\|\phi_N^{(e+\text{1})}-\phi_N^{(e)}\|_\text{2}$
    over the training epochs.}
    \label{fig:meta_stability}
    \vspace{2em}
\end{figure}

\textit{2) Theoretical Analysis of GML:}
In GML, Adam is employed to update the meta parameters $\phi_\mathbf{u}$, $\phi_\mathbf{W}$, 
and $\phi_\mathbf{N}$ in the epoch loop according to \eqref{update 1} \eqref{update 2} \eqref{update 3}. As illustrated in Fig.~\ref{fig:meta_stability}, the successive met parameter variations gradually decrease and become small after sufficient training epochs. This empirical observation suggests that the learned update rule enters a relatively stable regime. Accordingly, in the following analysis, we characterize the inner loop dynamics under a fixed learned update realization $\mathcal{H}=\{\phi_\mathbf{u}^\text{fixed},\phi_\mathbf{W}^\text{fixed},\phi_\mathbf{N}^\text{fixed}\}$.
Let \begin{equation}
{\cal L}(\boldsymbol{\beta}) \triangleq -R_S(\boldsymbol{\beta})
+ \lambda_1 {{\cal L}_1}(\boldsymbol{\beta})
+ \lambda_2 {{\cal L}_2}(\boldsymbol{\beta}),
\end{equation}
where $\boldsymbol{\beta}=\{ \mathbf{u}, \mathbf{W}, \mathbf{N} \}$ denotes
the optimization variables updated in the inner loop.

We first note that, under the transmit power constraint in \eqref{P1 constraintt 1},
the secrecy rate is upper bounded for any fixed channel
realization. Specifically, by the Cauchy-Schwarz inequality,
\begin{equation}
\left| \mathbf{h}_k^H \mathbf{w}_k \right|^2
\le
\left\| \mathbf{h}_k \right\|^2
\left\| \mathbf{w}_k \right\|^2 .
\end{equation}
Let
\begin{equation}
H_{\max} \triangleq \max_k \left\| \mathbf{h}_k \right\|^2 .
\end{equation}
Then, under $\text{tr}(\mathbf{R}_\text{X})\le P_\text{T}$, we have
\begin{equation}
\text{SINR}_k
\le
\frac{H_{\max} P_\text{T}}{\sigma^2}.
\end{equation}
Therefore,
\begin{equation}
R_\text{S}
\le
\sum_{k=1}^{K}\log_2\!\left(1+\text{SINR}_k\right)
\le
K\log_2\!\left(1+\frac{H_{\max}P_\text{T}}{\sigma^2}\right),
\end{equation}
which implies that $R_\text{S}$ is upper bounded for any fixed channel
realization. Since ${\cal L}_\text{1}$ and ${\cal L}_\text{2}$ are nonnegative by construction,
the loss function ${\cal L}(\boldsymbol{\beta})$ is lower bounded.

Next, consider the inner loop recursion
\begin{equation}
\boldsymbol{\beta}_{i+1}
=
\boldsymbol{\beta}_i + \mathbf{d}_i,
\qquad
\mathbf{d}_i
=
\mathcal{M}_{\mathcal{H}}\!\left(\nabla R_\text{S}(\boldsymbol{\beta}_i)\right),
\end{equation}
where $\mathcal{H}$ is fixed throughout the subsequent analysis, $\mathcal{M}_{\mathcal{H}}\left(\cdot \right)$ abstractly denotes the neural network operator parameterized by $\mathcal{H}$ .

\begin{assumption}\label{assumption1}
The loss function ${\cal L}(\boldsymbol{\beta})$ is continuously differentiable and $L_F$-smooth, where $L_F>\text{0}$ denotes the Lipschitz constant of the gradient of ${\cal L}(\boldsymbol{\beta})$, i.e.,
\begin{equation}
\left\|
\nabla {\cal L}(\boldsymbol{\beta}_1)
-
\nabla {\cal L}(\boldsymbol{\beta}_2)
\right\|
\le
L_F
\left\|
\boldsymbol{\beta}_1-\boldsymbol{\beta}_2
\right\|,
\quad
\forall \boldsymbol{\beta}_1,\boldsymbol{\beta}_2.
\end{equation}
\end{assumption}

To facilitate the subsequent analysis, we further impose the following sufficient condition on the learned inner loop step under the fixed realization $\mathcal{H}$.

\begin{assumption}\label{assumption2}
For the fixed learned update $\mathcal{M}_{\mathcal{H}}$, suppose that there exist constants $c_\text{1}>\text{0}$ and $c_\text{2}>\text{0}$ such that, for all inner-loop iterates, the induced update step $\mathbf{d}_i$ satisfies
\begin{equation}
\nabla {\cal L}(\boldsymbol{\beta}_i)^T \mathbf{d}_i
\le
-c_1 \left\| \nabla {\cal L}(\boldsymbol{\beta}_i) \right\|^2,
\end{equation}
and
\begin{equation}
\left\| \mathbf{d}_i \right\|
\le
c_2 \left\| \nabla {\cal L}(\boldsymbol{\beta}_i) \right\|.
\end{equation}
\end{assumption}

Assumption \ref{assumption2} serves as a sufficient condition for establishing the descent property of the induced inner loop dynamics under a fixed realization $\mathcal{H}$. It is introduced for the subsequent theoretical characterization, rather than to claim that the Adam-trained network globally guarantees such a property for all iterates.
\begin{proposition}
    \label{convengence_proof}
Under Assumption \ref{assumption1} and Assumption \ref{assumption2}, if
\begin{equation}
\delta
\triangleq
c_1-\frac{L_F}{2}c_2^2
>0,
\end{equation}
then the inner loop iterates satisfy
\begin{equation}
{\cal L}(\boldsymbol{\beta}_{i+1})
\le
{\cal L}(\boldsymbol{\beta}_{i})
-
\delta
\left\|
\nabla {\cal L}(\boldsymbol{\beta}_{i})
\right\|^2.
\end{equation}
 \end{proposition}
 
 \textit{Proof:} Please refer to Appendix \ref{appendixD}.$\hfill\blacksquare$

Consequently, for any $T \ge \text{1}$,
\begin{equation}
\sum_{i=1}^{T}
\left\|
\nabla {\cal L}(\boldsymbol{\beta}_i)
\right\|^2
\le
\frac{
{\cal L}(\boldsymbol{\beta}_1)-{\cal L}(\boldsymbol{\beta}_{T+1})
}{\delta}.
\end{equation}
Since ${\cal L}(\boldsymbol{\beta})$ is lower bounded, letting $T\to\infty$ yields
\begin{equation}
\sum_{i=1}^{\infty}
\left\|
\nabla {\cal L}(\boldsymbol{\beta}_i)
\right\|^2
< \infty.
\end{equation}
Because each term in the above series is nonnegative, it follows that
\begin{equation}
\lim_{i\rightarrow\infty}
\left\|
\nabla {\cal L}(\boldsymbol{\beta}_i)
\right\|
=0.
\end{equation}
Therefore, if the sequence $\{\boldsymbol{\beta}_i\}$ admits an accumulation point, any such accumulation point is a first-order stationary point of ${\cal L}$.

\section{Numerical Results}\label{sec4}
In this section, we present numerical results to evaluate the
performance of the proposed algorithm in the MA-enabled secure ISAC system. The simulation parameters are detailed in {\bf{Table}} \ref{tab:table2}. The simulations primarily evaluate communication and sensing performance, and a comparison is provided against various baseline algorithms. These baseline algorithms are enumerated as follows:
\begin{itemize}
    \item {\bf{Baseline 1}} (ML): Meta learning (ML) is a simplified
version of GML which removes the  gradient-based input mechanism.
    \item {\bf{Baseline 2}} (PGA): Projected gradient ascent (PGA) \cite{MASecure} is employed to optimize the MA part, while variables $\mathbf{W}$ and $\mathbf{N}$ are still optimized using GML.
    \item {\bf{Baseline 3}} (FPA): The dual-functional BS employs an FPA array, while variables $\mathbf{W}$ and $\mathbf{N}$ are still optimized using GML.
    \item {\bf{Baseline 4}} (RA): The dual-functional BS employs a randomly distributed antenna array, while variables $\mathbf{W}$ and $\mathbf{N}$ are still optimized using GML.
    \item {\bf{Baseline 5}} (AO): Alternately optimizes antenna positioning, transmit beamforming, and artificial noise by solving first order convex approximated subproblems with CVX.
    \item {\bf{Baseline 6}} (MVPSO): Multi-velocity particle swarm optimization (MVPSO) \cite{MVPSO} is used for antenna positioning search and CVX based convex approximation is used to update transmit beamforming and artificial noise alternately.
\end{itemize}

\begin{table}[t] 
\vspace{3.5em}
    \centering 
    \caption{Simulation Parameters}
    \label{tab:table2} 
    \begin{tabular}{lc}
        \toprule
        Parameters  & Values   \\
        \midrule
        \multicolumn{2}{l}{\textbf{\textit{System Parameters}}} \\
        Carrier wavelength $\lambda$  & 0.01$\text{m}$ \\
         Minimum distance between any two MA elements $d_\text{min}$ & 0.5$\lambda$  \\
        Path loss exponent   $o$ & 2.5  \\
        Propagation loss of eavesdropper  $\beta$ & $\text{2.7}\times\text{10}^\text{-4}$  \\
        Noise power  $\sigma^\text{2}$ & 1$\times\text{10}^\text{-7}\text{mW}$  \\
        Beam width  $\Delta$ & 9$^\circ$  \\
        The threshold of sense quality  $\xi$ & 0.5\\
        \addlinespace 
        \multicolumn{2}{l}{\textbf{\textit{Algorithm Parameters}}} \\
        Learning rate of \textit{MN}  $\alpha_{\mathbf{u}}$ & 3$\times\text{10}^{\text{-3}}$  \\
        Learning rate of \textit{BN}  $\alpha_{\mathbf{W}}$ & 1$\times\text{10}^{\text{-3}}$  \\
        Learning rate of \textit{AN}  $\alpha_{\mathbf{N}}$ & 3$\times \text{10}^{\text{-3}}$  \\
        Number of inner loop  $N_\text{i}$ & 1  \\
        Number of outer loop  $N_\text{o}$ & 1  \\
        Number of epoch loop  $N_\text{e}$ & 2000  \\
        Number of channel in simulation  $N_\text{a}$ & 10  \\
        Update interval of $\mathbf{R}_\text{d}$ &100\\
        Update interval of $\mathbf{W,N}$ & 5\\
        \bottomrule
    \end{tabular}
\end{table}
    
    
	
\subsection{Communication Performance}
In this subsection, we evaluate the communication performance of the
proposed algorithm against baselines. All the simulation curves have been averaged over $N_\text{a}$ independent channel realizations.

\begin{figure}[t]  
    \centering
\includegraphics[width=0.85\linewidth]{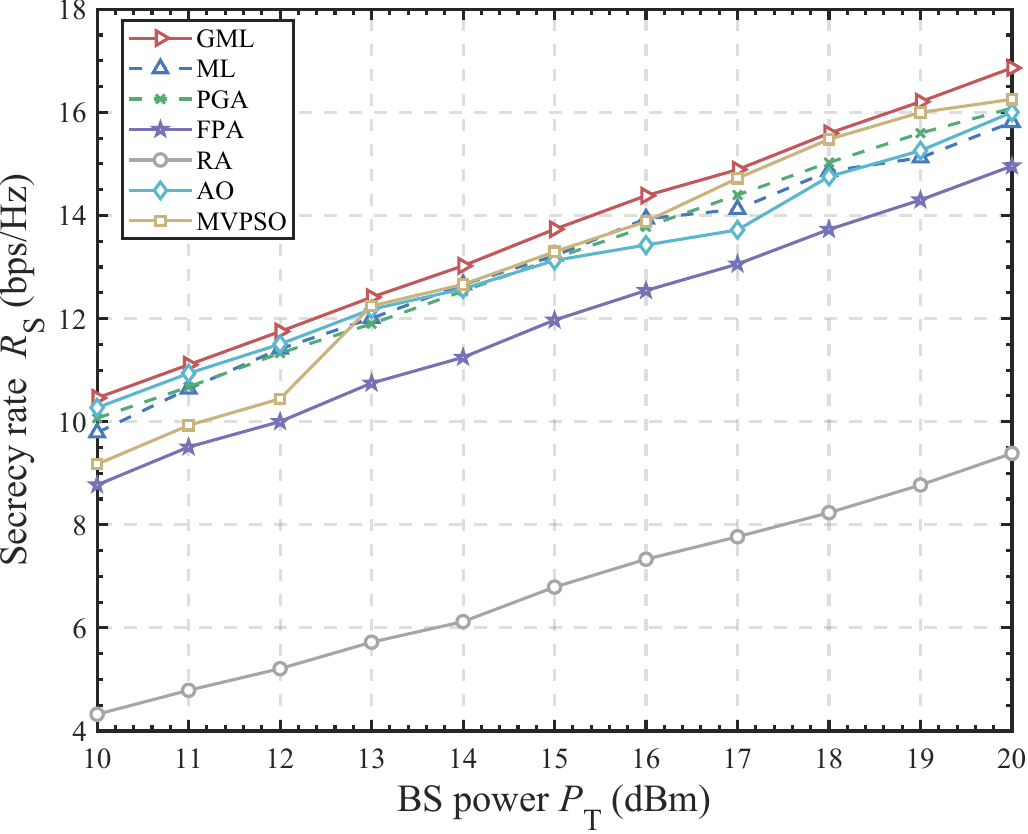}
    \caption{\ Secrecy rate $R_\text{S} $ versus the BS power $P_\text{T}$.}
    \label{fig:4_main}
    \vspace{2em}
\end{figure}

We first present the optimization performance and comparison results of the algorithms. Fig.~\ref{fig:4_main} illustrates the optimization performance versus the power of the dual-functional BS, with the number of antennas $M$ fixed at 7 and the number of legitimate users $K$ fixed at 2. The two legitimate users are located at azimuths of $\pm \text{30}^{\circ}$, while the eavesdropper is located at an azimuth of $\text{15}^{\circ}$. As observed in the Fig.~\ref{fig:4_main}, the  optimization results of all algorithms improve  with the increase in BS power $P_\text{T}$. It is evident that RA performs the worst among the baseline algorithms due to its random antenna distribution. In contrast, the GML, ML and PGA algorithms, by optimizing the MA elements, exploit their spatial DoFs which are superior to those of the FPA array. This indicates that optimizing the antenna positions based on the channel scenario can significantly enhance system performance. 
Among the five MA optimization algorithms, GML exhibits the best performance, as it leverages both high-dimensional gradient information input and the superiority of meta learning. It is worth noting that the proposed GML achieves  $\text{4.83}\%$ and $\text{6.67}\%$ higher performance than PGA and ML at $P_\text{T}=\text{20 dBm}$. Correspondingly, ML lacks the high-dimensional gradient-baesd input, while the optimizer for MA in PGA remains fixed throughout the optimization process. It is not visually intuitive from this figure which of ML and PGA is superior. This essentially reflects the relative advantages of high-dimensional gradient-based input versus meta learning under the data distributions of different scenarios. For both AO and MVPSO, the AO method relies on local convex approximations and trust-region update mechanisms, which may limit its ability to escape local stationary points. In contrast, MVPSO adopts a derivative-free antenna positioning search strategy, and its performance is affected by the finite number of particles and the alternating update mechanism.\\
\vspace{0em}

\begin{figure}[t]  
    \centering
\includegraphics[width=0.85\linewidth]{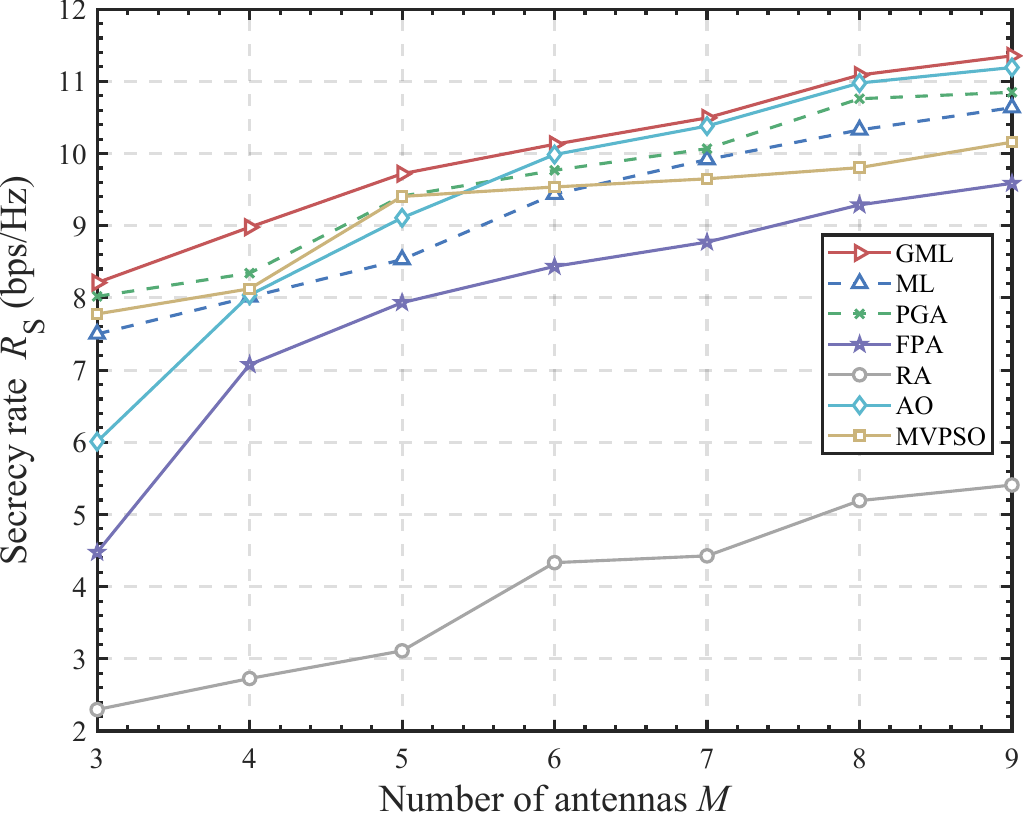}
    \caption{\ Secrecy rate $R_\text{S} $ versus the number of antennas $M$.}
    \label{fig:6_main}
    \vspace{2em}
\end{figure}
\vspace{-1em}
Without changing the azimuths of the legitimate users and the eavesdropper, Fig.~\ref{fig:6_main} illustrates the algorithm's optimization performance versus the number of dual-functional BS antennas, when the BS power $P_\text{T}$ is fixed at 10 dBm. It is evident that system performance improves as the number of antennas increases due to the enhanced array gain and spatial DoFs afforded by a larger array. Also it can be observed that, benefiting from spatial DoFs, the MA performance remains significantly superior to that of FPA and RA. Furthermore, the superiority of PGA over ML demonstrates that, under the data distribution of this scenario, the gain from the high-dimensional information introduced by gradients outweighs the advantages brought by meta learning. Finally, the fact that GML outperforms PGA once again validates the rationality and effectiveness of the mapping strategy for MA coordinates adopted in this paper. The performance gap
between GML and MVPSO becomes more evident as $M$ increases, because the
dimension of the antenna positioning search space grows with $M$, making the
particle based search more likely to suffer from insufficient exploration under
a limited particle budget. By contrast, GML adaptively learns the update direction from the
gradient of each optimization block, which enables it to make more effective
use of the additional spatial degrees of freedom.

\begin{figure}[t]  
    \centering
\includegraphics[width=0.85\linewidth]{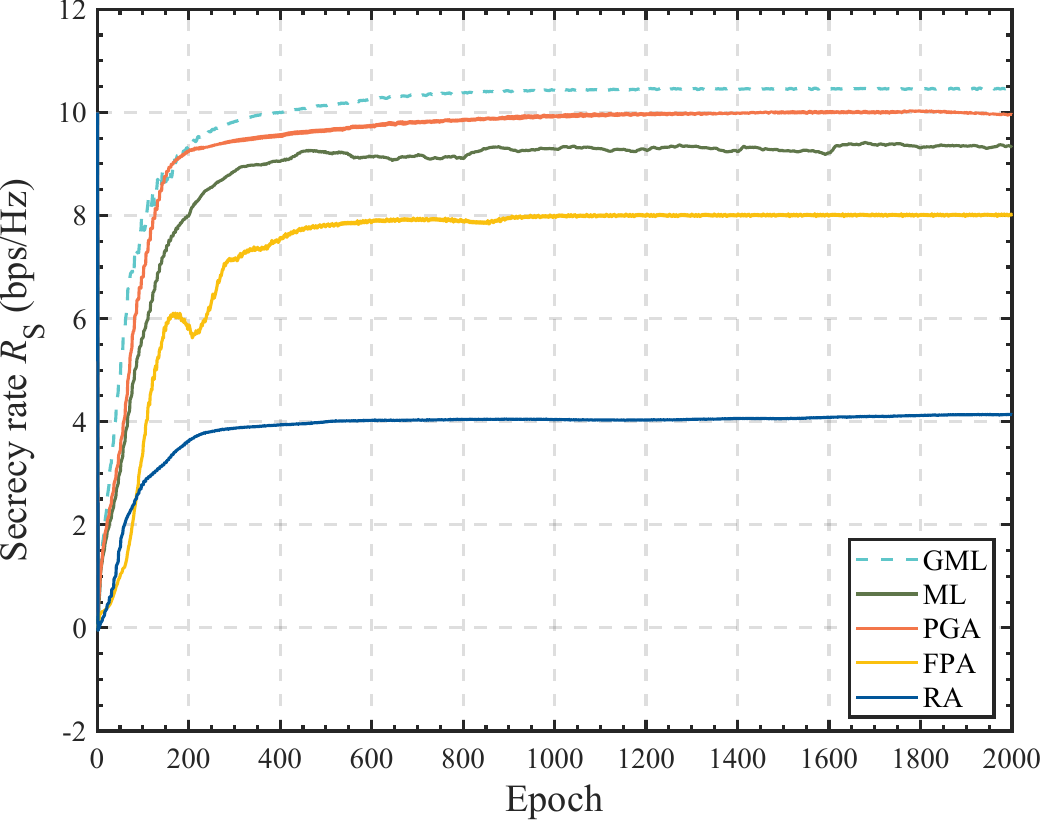}
    \caption{\ Secrecy rate $R_\text{S} $ versus the epoch.}
    \label{fig:5_main}
    \vspace{2em}
\end{figure}
Then, Fig.~\ref{fig:5_main} shows the convergence behavior of the proposed algorithms under the setting: $P_\text{T}=\text{10 dBm}, M=\text{7}$ and the same azimuths of the legitimate users and the eavesdropper as Fig.~\ref{fig:4_main}. It can be observed that GML, PGA, and FPA all demonstrate good convergence properties. In contrast, ML exhibits poor convergence because it lacks high-dimensional gradient information input, which leads to instability during the joint optimization process of the multiple variables. Besides, it is worth noting that GML
displays a steeper rise than PGA during the initial epochs. This is due to the unconstrained transformation applied to $\mathbf{p}$, thus the solution space of the GML optimization algorithm is open and continuous. Consequently, from the perspective of the solution space's geometry and topology, the GML  can perform an unimpeded and efficient search along the optimal descent direction, thus making GML easier to approach the optima and
achieve higher performance than PGA. The findings presented from Fig.~\ref{fig:4_main} to Fig.~\ref{fig:5_main}, covering both communication performance and convergence performance, collectively validate the rationality and effectiveness of the processing strategy employed for $\mathbf{p}$. 

\vspace{0em}
\vspace{0em}
\begin{figure}[t]  
    \centering
\includegraphics[width=0.85\linewidth]{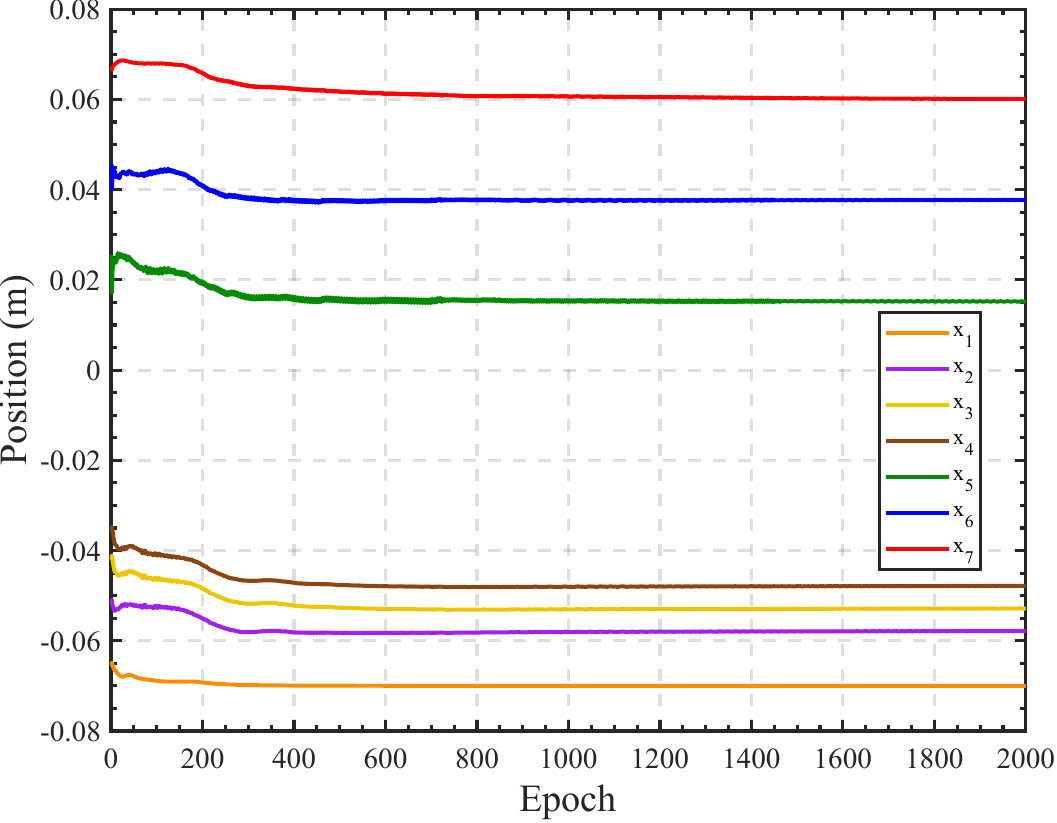}
    \caption{\ MA positions versus the epoch.}
    \label{fig:7_main}
    \vspace{2em}
\end{figure}
Next, Fig.~\ref{fig:7_main} demonstrates the convergence results of antenna positions in the GML  under the setting: $P_\text{T}=\text{10 dBm}, M=\text{7}$. It can be seen that the result essentially reaches convergence after 400 epochs, which further demonstrates the convergence performance of the GML optimization algorithm. The converged antenna positions, as determined by the GML , form a channel-adaptive array geometry specifically optimized to the current user distribution and propagation environment, rather than a predefined uniform layout.

\begin{figure}[t]  
    \centering
\includegraphics[width=0.85\linewidth]{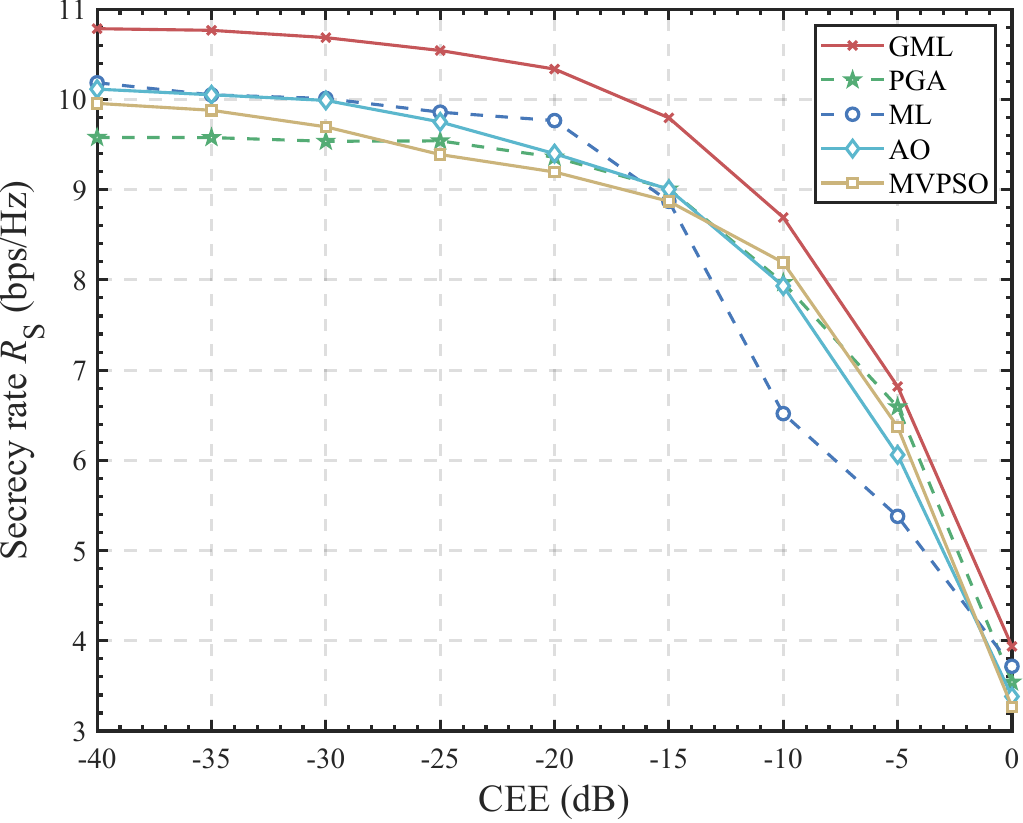}
    \caption{\ Secrecy rate $R_\text{S} $ versus the CEE.}
    \label{fig:rate_vs_CEE}
    \vspace{2em}
\end{figure}

Considering the difficulty of acquiring perfect channel state information (CSI), we define $\mathbf{h}$ as the perfect channel and $\hat{\mathbf{h}}$ as the estimated channel to account for the channel estimation error (CEE). Then, the CEE can be expressed as follows:
\begin{equation}
\text{CEE} = 10 \log_{10} \left( 
\frac{\mathbb{E}\left[\lVert \mathbf{h} - \hat{\mathbf{h}} \rVert_2^2 \right]}
{\mathbb{E}\left[\lVert \mathbf{h} \rVert_2^2 \right]}
\right),\label{CEE}
\end{equation}
where a smaller CEE value indicates 
more accurate channel estimation. 
For the imperfect CSI case, the estimation 
error $\mathbf{z}=\mathbf{h}-\hat{\mathbf{h}}$ is modeled as a 
zero-mean additive circularly symmetric complex white Gaussian random
 variable \cite{Meta4}.
As illustrated in Fig.~\ref{fig:rate_vs_CEE}, we investigate the variation trend of the secrecy rate versus the CEE under a fixed transmit power $P_\text{T}$ set to $\text{10 dBm}$. First, it is observed that as the CEE increases, the secrecy rates of all algorithms exhibit a downward trend. However, GML consistently demonstrates superior performance under imperfect CSI across varying CEE levels. This implies that GML relaxes the strict requirement for near-perfect channel information, thereby reducing the overall system overhead. In other words, to achieve an identical secrecy rate, GML can tolerate a higher degree of channel uncertainty. Fundamentally, this advantage can be attributed to the gradient-input and meta learning mechanisms of GML.

\begin{figure}[t]  
    \centering
\includegraphics[width=0.9\linewidth]{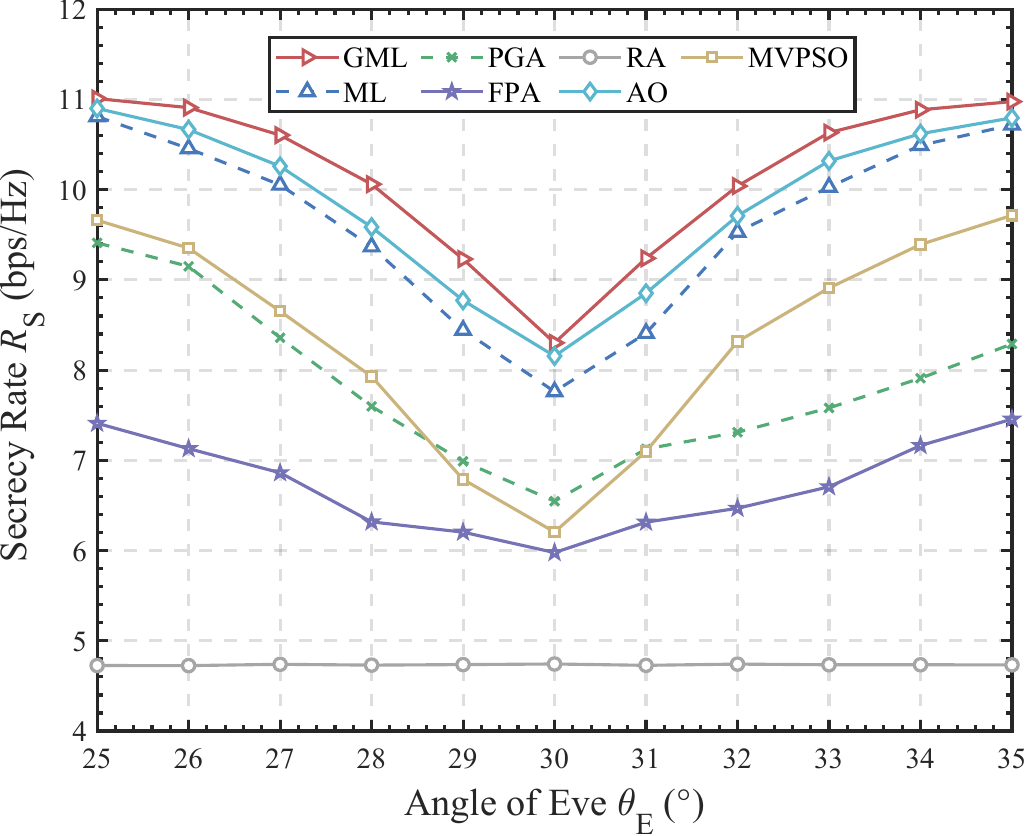}
    \caption{\ Secrecy rate $R_\text{S} $ versus the Eve angle $\theta_\text{E}$.}
    \label{fig:rate_vs_angle}
    \vspace{2em}
\end{figure}
Fig.~\ref{fig:rate_vs_angle} depicts the secrecy rate versus the eavesdropper
angle $\theta_{\rm E}$. The secrecy rate exhibits a clear valley around
$\theta_{\rm E}=\text{30}^\circ$. This is because the eavesdropper becomes angularly
close to one legitimate user, which increases the channel correlation between
the legitimate and wiretap links and makes secure spatial separation more
difficult. For the proposed GML, the secrecy rate decreases from about
$\text{11.0}$ bps/Hz at $\theta_{\rm E}=\text{25}^\circ$ to about $\text{8.3}$ bps/Hz at
$\theta_{\rm E}=\text{30}^\circ$, and then increases again to about $\text{11.0}$ bps/Hz
at $\theta_{\rm E}=\text{35}^\circ$. At the most challenging point
$\theta_{\rm E}=\text{30}^\circ$, GML still outperforms other baselines.
The almost flat and low performance of RA indicates that random antenna
deployment cannot adapt to the angular variation of the eavesdropper. FPA is
also limited by its fixed geometry. MVPSO and AO can exploit MA
degrees of freedom, but MVPSO is affected by the multimodal particle-search
landscape and AO is restricted by local convexification. GML achieves the
highest secrecy rate over the entire angular range because it jointly adapts
the antenna positioning, beamforming, and artificial noise according to
the instantaneous gradient information.
\vspace{0em}
\begin{figure}[H]  
    \centering
\includegraphics[width=0.9\linewidth]{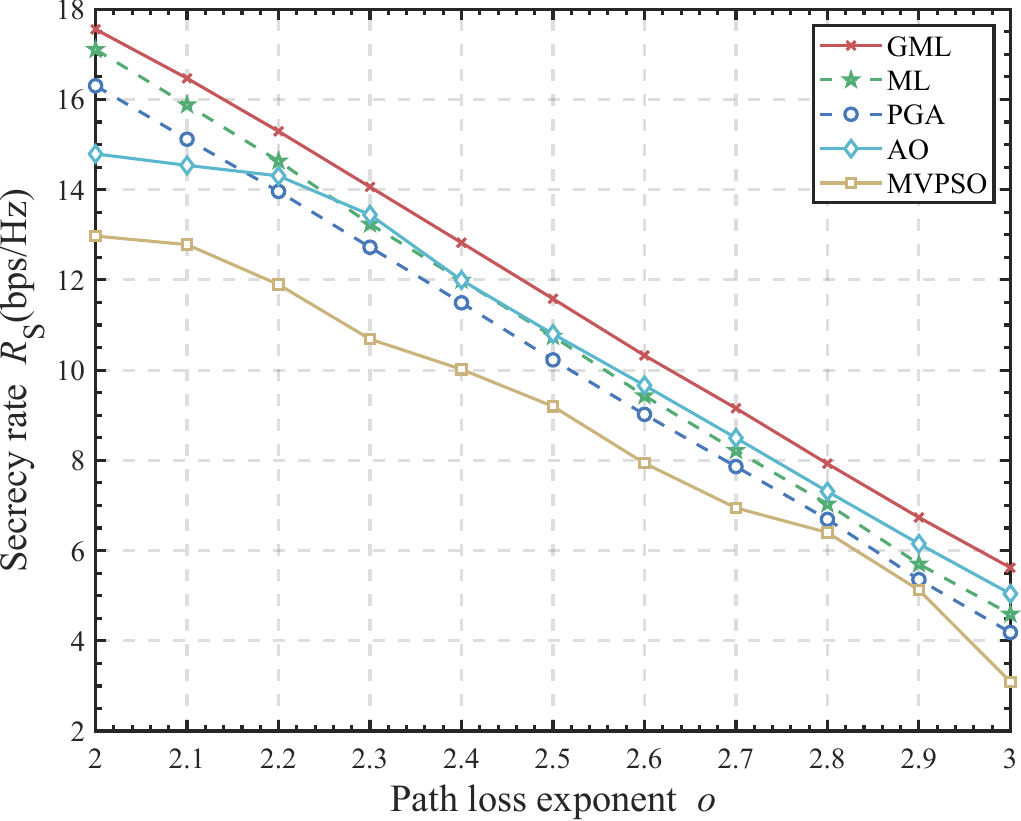}
    \caption{\ Secrecy rate $R_\text{S} $ versus the path loss exponent $o$.}
    \label{fig:fading_factor}
    \vspace{2em}
\end{figure}
\vspace{0em}

Fig.~\ref{fig:fading_factor} shows the secrecy rate $R_{\text{S}}$ versus the path loss exponent $o$. As $o$ increases from 2 to 3, the secrecy rates of all schemes decrease significantly due to the more severe large scale propagation attenuation, which weakens the received signal power and reduces the effectiveness of secure beamforming and artificial noise design. The proposed GML consistently achieves the best performance over the whole range. Even under the severe path loss condition of $o=\text{3}$, GML still maintains the highest secrecy rate of about 5.6 bps/Hz. This advantage mainly comes from its ability to exploit instantaneous gradient information and learn adaptive update rules for the joint optimization of antenna positioning, beamforming, and artificial noise. In contrast, AO is limited by its local convex approximations, while MVPSO suffers from inefficient particle based search in the high dimensional non-convex space. Moreover, ML and PGA lack either gradient input mechanism or meta learning approach, which restricts their performance under severe path loss conditions.

\subsection{Sensing Performance}
In this subsection, we primarily evaluate the beampattern performance of GML under various system parameters, including MA region sizes, the number of MA elements, and different distributions of legitimate users and eavesdroppers. 
\vspace{0em}
\begin{figure}[H]  
    \centering
\includegraphics[width=0.85\linewidth]{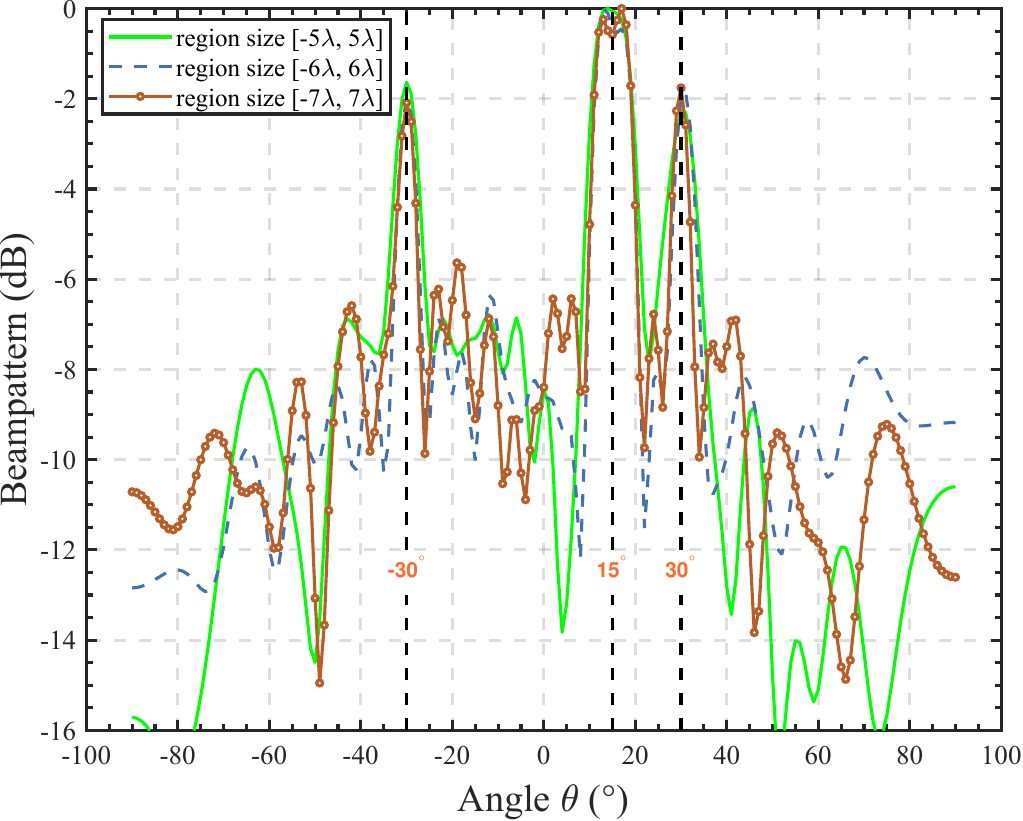}
    \caption{\ Beampattern versus the region size of MA.}
    \label{fig:region size}
    \vspace{1.2em}
\end{figure}
\vspace{0.5em}
Fig.~\ref{fig:region size} illustrates the beampattern results under different MA region sizes, with $M$ fixed at 13 and $P_\text{T}$ at 10 dBm. The legitimate users are located at $\pm \text{30}^{\circ}$, and the eavesdropper is positioned at $\text{15}^{\circ}$. It can be observed that for all considered  region sizes, GML successfully generates precise high-gain beams aligned with the target direction. This validates that by flexibly adjusting the antenna positions within a continuous linear region, the MA can effectively reconstruct the channel to support beamforming. Moreover, the performance exhibits robustness against the variation of the moving region size. Even with a limited region sizes of $L=\text{5}\lambda$, the MA is capable of exploiting sufficient DoFs along the linear dimension to form distinct main lobes comparable to those achieved with a larger region (e.g., $L=\text{7}\lambda$). This suggests that GML can achieve satisfactory interference suppression and beamforming gain within a compact linear space, making it highly practical for deployment scenarios with dimensional constraints.

\vspace{0em}
\begin{figure}[H]  
    \centering
\includegraphics[width=0.85\linewidth]{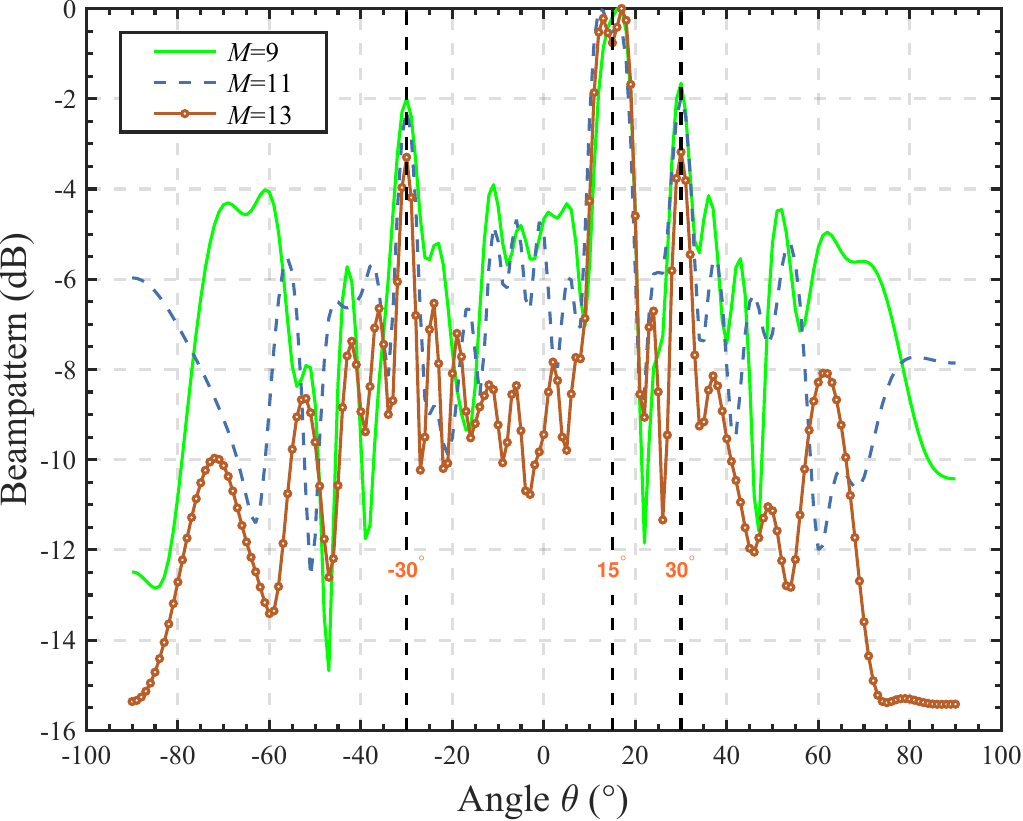}
    \caption{\ Beampattern versus the number of antennas $M$.}
    \label{fig:wave_vs_antenna}
    \vspace{1.2em}
\end{figure}
\vspace{0.5em}
Fig.~\ref{fig:wave_vs_antenna} illustrates the beampattern with varying numbers of antenna elements $M \in \{\text{9}, \text{11}, \text{13}\}$, within a fixed transmit region $[-\text{7}\lambda, \text{7}\lambda]$. As shown in Fig.~\ref{fig:wave_vs_antenna}, GML accurately forms high-gain main lobes towards all communication and sensing directions regardless of the value of $M$. It can be observed that while all three configurations successfully form accurate beams towards the target directions, increasing the number of MA elements yields a significant improvement in sidelobe suppression. This trend demonstrates that a larger array scale provides higher spatial resolution and DoFs. Consequently, the MA system can concentrate energy more sharply on the intended directions (both for communication and sensing) while minimizing interference in other directions, thereby enhancing the overall security and spectral efficiency of the system.

\vspace{0em}
\begin{figure}[H]  
    \centering
\includegraphics[width=0.85\linewidth]{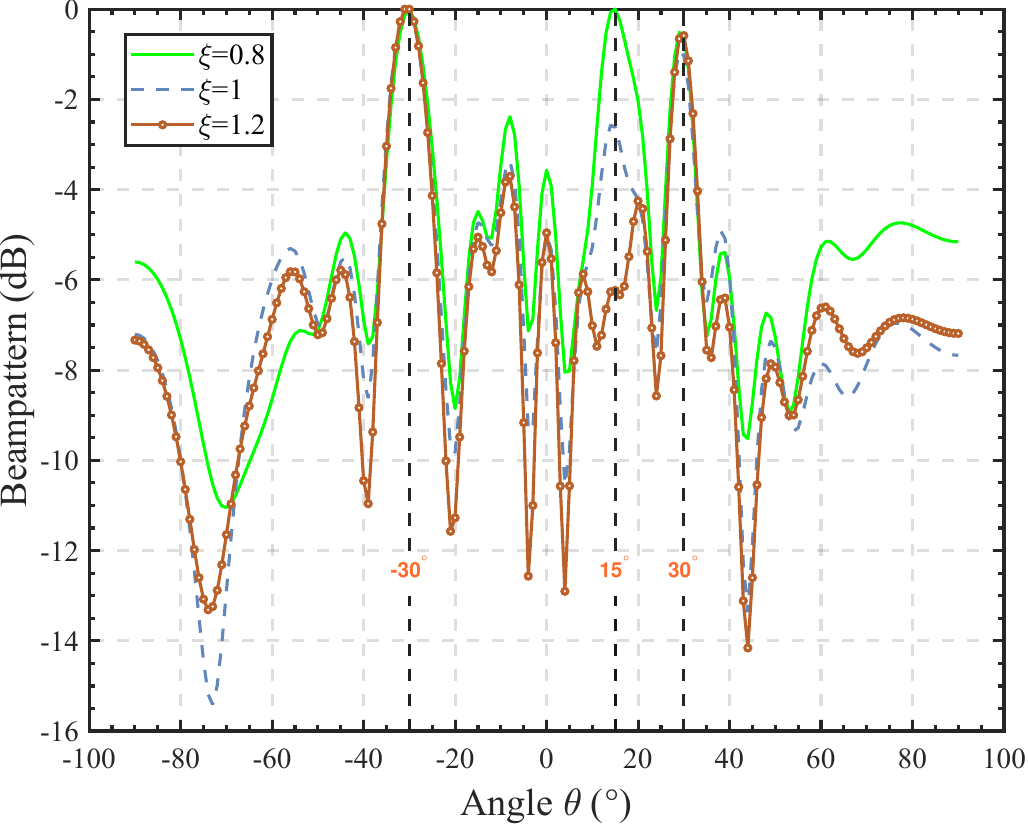}
    \caption{\ Beampattern versus the sensing threshold $\xi$.}
    \label{fig:wave_vs_yuzhi}
    \vspace{1.2em}
\end{figure}
\vspace{0.5em}
Fig.~\ref{fig:wave_vs_yuzhi} shows the normalized beampattern generated by GML under
different sensing thresholds $\xi$. The three vertical dashed lines correspond
to the  angular directions, i.e., $-\text{30}^\circ$, $\text{15}^\circ$, and
$\text{30}^\circ$. It can be observed that GML forms clear beams around these
legitimate directions for all considered $\xi$, which means the communication beam design is strictly independent of the sensing threshold setting.  In addition, several deep nulls below $-\text{12}$ dB can be observed, which indicates that the optimized transmit covariance
can simultaneously provide directional energy focusing and interference
suppression. When $\xi$ decreases from $\text{1.2}$ to $\text{0.8}$, the sensing constraint
becomes stricter, and the response around the sensing direction
$\text{15}^\circ$ increases from roughly $-\text{6}$ dB to nearly $\text{0}$ dB. This confirms that
a smaller $\xi$ forces the optimized covariance to better match the desired
sensing beampattern. On the other hand, a larger $\xi$ relaxes the sensing
requirement and gives GML more freedom to allocate transmit covariance for
secrecy rate maximization, which leads to different sidelobe shapes. These
results verify that GML can flexibly balance the sensing beampattern constraint
and the secure communication objective.

\begin{figure}[H]  
    \centering
\includegraphics[width=0.85\linewidth]{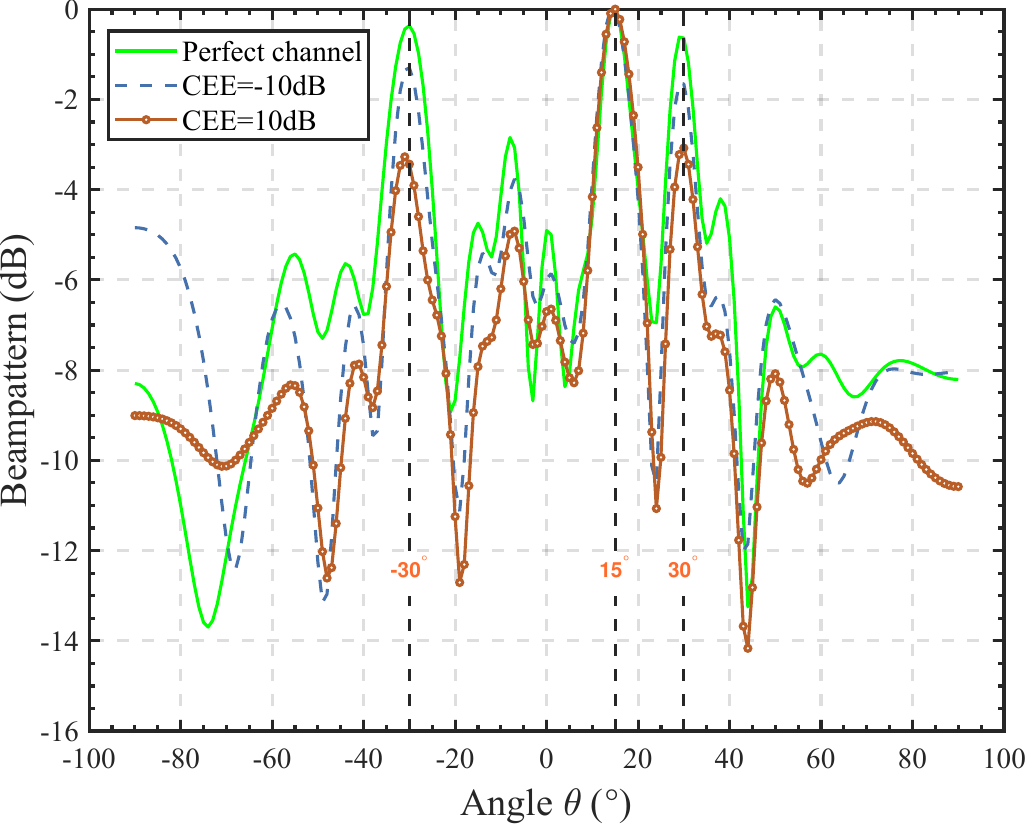}
    \caption{\ Beampattern versus the CEE.}
    \label{fig:wave_vs_CEE}
    \vspace{1.2em}
\end{figure}

Fig.~\ref{fig:wave_vs_CEE} illustrates the normalized beampattern of GML under
different CEE levels. With perfect channel information, GML generates
well-aligned main beams at the target and legitimate directions, where the sensing
direction around $\text{15}^\circ$ reaches the normalized peak and the two
communication directions around $\pm \text{30}^\circ$ remain within
approximately $\text{1}$ dB of the peak. When $\text{CEE}=-\text{10}$ dB, the sensing directions are still preserved, but the gains around $\pm \text{30}^\circ$ decrease
to roughly $-\text{1.5}$ dB to $-\text{2}$ dB. Under the more severe case
$\text{CEE}=\text{10}$ dB, the gains around $\pm \text{30}^\circ$ further decrease to
about $-\text{3}$ dB.
Nevertheless, the main lobe around the sensing direction $\text{15}^\circ$ is still
maintained close to $\text{0}$ dB, showing that the proposed GML can preserve the
essential sensing functionality even under imperfect channel due to \eqref{P2 constraint 2}. The beampattern distortion caused by CEE is consistent with the secrecy rate degradation observed in Fig.~\ref{fig:rate_vs_CEE}. Specifically, although channel uncertainty weakens the beamforming accuracy towards legitimate users, GML still maintains robust directional focusing due to its joint optimization of antenna positioning, beamforming, and artificial noise.
\vspace{0em}
\vspace{0.5em}

\begin{figure*}[ht]
    \centering
    
    \subfigure[Legitimate users at $\pm \text{30}^{\circ}$ and eavesdropper at $\text{15}^{\circ}$.]{
        \includegraphics[width=0.31\textwidth]{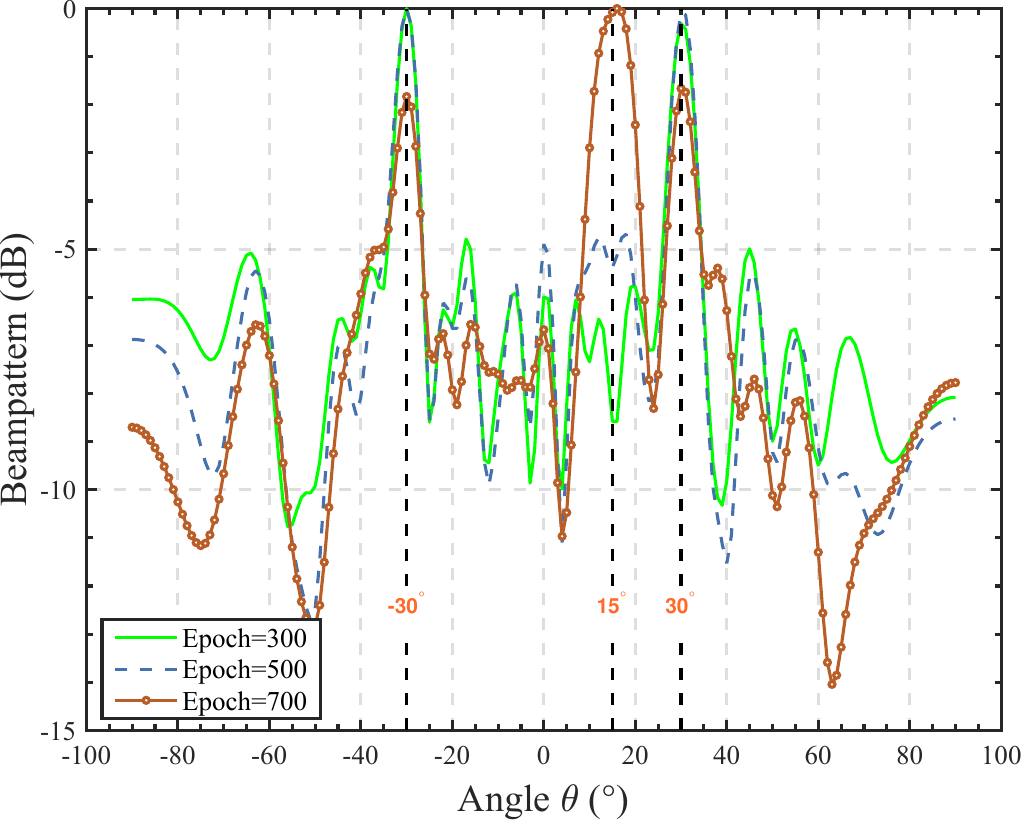}
        \label{fig:wave_epoch1}
    }
    \hfill
    \subfigure[Legitimate users at $\pm \text{30}^{\circ},\text{0}^{\circ}$ and eavesdropper at $\text{15}^{\circ}$.]{
        \includegraphics[width=0.31\textwidth]{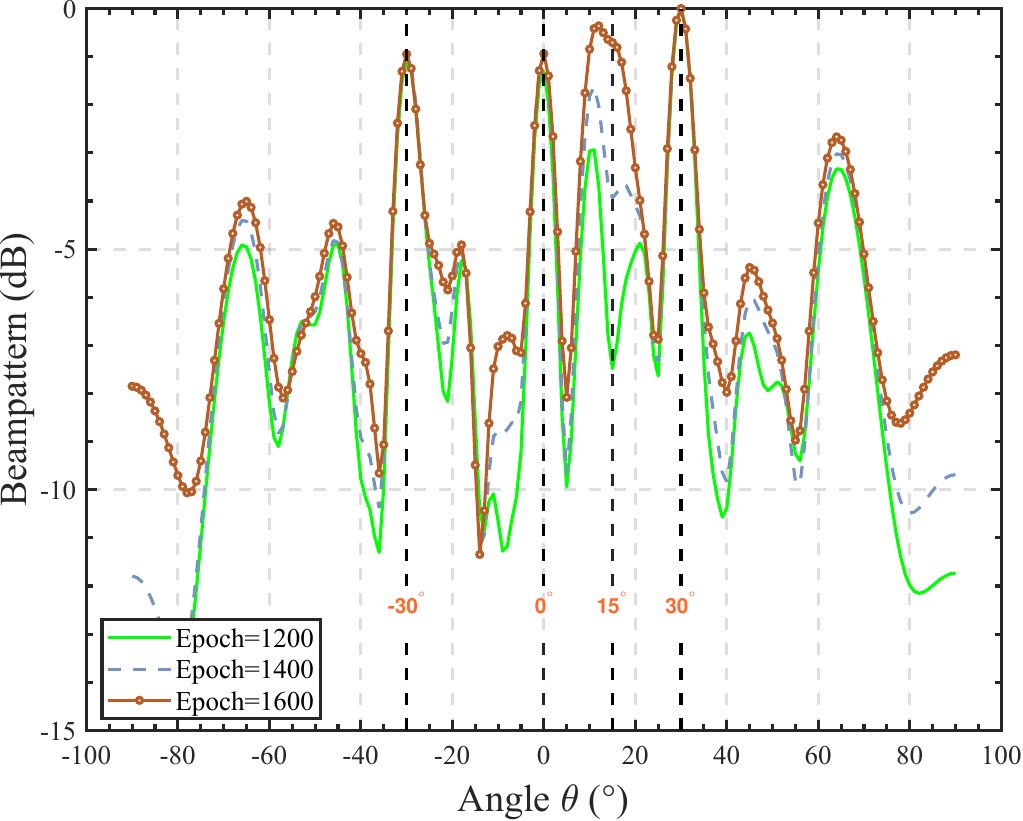}
        \label{fig:wave_epoch2}
    }
    \hfill
    \subfigure[Legitimate users at $\pm \text{30}^{\circ},\pm\text{60}^{\circ}$ and eavesdropper at $\text{15}^{\circ}$.]{
        \includegraphics[width=0.31\textwidth]{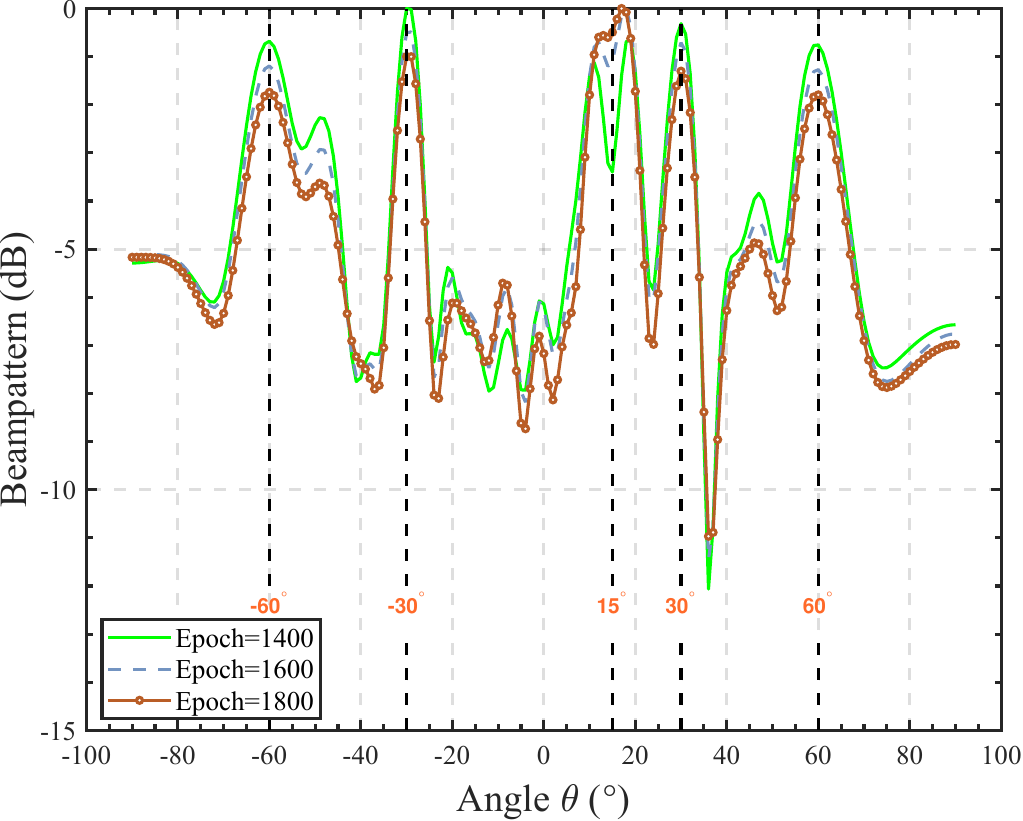}
        \label{fig:wave_epoch3}
    }

    \caption{Beampattern versus epoch under different legitimate users and eavesdropper distributions.}
    \vspace{1em}
    \label{fig:wave_epoch}
\end{figure*}
Fig.~\ref{fig:wave_epoch} depicts the beampattern evolution of the GML  at different epochs under various legitimate user distributions. It is noteworthy that all three subplots Fig.~\ref{fig:wave_epoch1}, Fig.~\ref{fig:wave_epoch2} and Fig.~\ref{fig:wave_epoch3} exhibit a consistent trend: as the epochs progress and the algorithm iterates, the joint waveform gradually approaches the direction of the sensing target (the eavesdropper), while maintaining stable beam alignment for the legitimate users. Furthermore, a detailed observation of each subplot reveals that as the algorithm iterates, the beam gain at the legitimate users experiences a slight decline, whereas the gain towards the sensing target sees a minor increase. This intrinsically reflects the trade-off between communication performance and sensing performance.
    


\section{Conclusion}\label{sec5}
This paper investigates a joint optimization algorithm for 
antenna positioning and beamforming in MA-enabled secure ISAC systems. 
First, we establish the system model for the MA-enabled secure ISAC 
system. By comprehensively considering both communication and sensing 
metrics, an optimization problem is formulated with the objective 
of maximizing the secrecy rate. 
Subsequently, we propose the GML optimization algorithm 
incorporated with various constraint handling strategies. 
This algorithmic framework is applied to the joint design 
of antenna positioning, beamforming, and artificial noise in 
the MA-enabled secure ISAC system. Finally, numerical simulations 
are conducted to compare the proposed algorithm with several 
baseline schemes. The experimental results verify the rationality 
of the adopted constraint handling strategies and the effectiveness o
f the proposed algorithm. The proposed algorithm 
offers new insights for MA-enabled secure ISAC systems. 
The proposed framework will be extended to more MA secure ISAC designs in future work. 
Specifically, explicit echo observation models, 
Cramér-Rao bound constrained sensing metrics, and adaptive penalty-based GML 
frameworks will be jointly investigated within this extended framework.
    \begin{appendices}

    \section{Proof of Lemma \ref{Lemma1}}
    \label{appendixA}
      \begin{lemma}
      \label{Lemma4}
      \textit{The gradient in $\text{P}_a$ and $\text{P}_b$ exist following equation} ${\nabla _{\bf{u}}}Z = {\bf{J}}_f^{\rm{T}} \cdot {\nabla _{\bf{p}}}Z,$ \textit{where ${\nabla _{\bf{u}}}Z\in \mathbb R^{M+\text{1}}$ denotes the gradient of $Z$ with respect to $\bf{u}$, ${\nabla _{\bf{p}}}Z\in \mathbb R^{M}$ denotes the gradient of $Z$ respect to $\bf{p}$, ${\bf{J}}_f\in \mathbb{R}^{M\times (M+\text{1})}$ denotes the Jacobian matrix of the vector function ${f}\left(  \cdot  \right)$, ${\bf{J}}_f(i,j)=\frac{{\partial {p_i}}}{{\partial {u_j}}}$.}
    \end{lemma}
    
    \textit{Proof.} According to the definition total differential
        \begin{equation}
        \begin{aligned}
        dZ &= \frac{{\partial Z}}{{\partial {p_1}}} \partial {p_1} + ... + \frac{{\partial Z}}{{\partial {p_N}}} \partial {p_N} \\&= \left[ {\frac{{\partial Z}}{{\partial {p_1}}},...,\frac{{\partial Z}}{{\partial {p_N}}}} \right] {\left[ {\partial {p_1},...,\partial {p_N}} \right]^{\rm{T}}},
        \end{aligned}
          \label{total differential1}
        \end{equation}
        denote ${\nabla _{\bf{p}}}Z = {\left[ {\frac{{\partial Z}}{{\partial {p_\text{1}}}},...,\frac{{\partial Z}}{{\partial {p_N}}}} \right]^{\rm{T}}}$, $d{\bf{p}} = {\left[ {\partial {p_\text{1}},...,\partial {p_N}} \right]^{\rm{T}}}$. Then it can be given as 
        \begin{equation}
        dZ = \nabla _{\bf{p}}^{\rm{T}}Z \cdot d{\bf{p}}.
        \label{proof1}
        \end{equation}
        Similarly, 
        \begin{equation}
        dZ = \nabla _{\bf{u}}^{\rm{T}}Z \cdot d{\bf{u}},
        \label{proof2}
        \end{equation}
         \begin{equation}
        d{\bf{p}} = {{\bf{J}}_f} \cdot d{\bf{u}}.
        \label{proof3}
        \end{equation}
        Combining \eqref{proof1}-\eqref{proof3}, it can be deduced that $\nabla _{\bf{u}}^{\rm{T}}Z \cdot d{\bf{u}} = \nabla _{\bf{p}}^{\rm{T}}Z \cdot d{\bf{p}} = \nabla _{\bf{p}}^{\rm{T}}Z \cdot \left( {{{\bf{J}}_f} \cdot d{\bf{u}}} \right) = \left( {\nabla _{\bf{p}}^{\rm{T}}Z \cdot {{\bf{J}}_f}} \right)  d{\bf{u}}$, that is,
        \begin{equation}
       \nabla _{\bf{u}}^{\rm{T}}Z = \nabla _{\bf{p}}^{\rm{T}}Z \cdot {{\bf{J}}_f},
        \end{equation}
        taking the transpose of both sides yields
        \begin{equation}
            {\nabla _{\bf{u}}}Z = {\bf{J}}_f^{\rm{T}} \cdot {\nabla _{\bf{p}}}Z.
        \end{equation}
        
        \begin{lemma}
        \label{Lemma5}
        \textit{Denote two matrix  $\mathbf{A}$ and $\mathbf{C}$, ${\mathbf{A}} \in {\mathbb{R}^{b \times a}}$, ${\mathbf{C}} \in {\mathbb{R}^{c \times b}}$, it follows that }${\rm{Rank}}\left( {{\bf{CA}}} \right) = {\rm{Rank}}\left( {\bf{A}} \right) - \text{dim} \left( {{\rm{Null}}\left( {\bf{C}} \right) \cap {\rm{Image}}\left( {\bf{A}} \right)} \right)$.
        \end{lemma}
        
         \textit{Proof.} According to the rank-nullity theorem, for a linear transformation $\bm{\phi}$,  it holds that
        \begin{equation}
        \text{dim} \left( {{\text{Domain}}} \right) = \text{dim} \left( {{\text{Image}}\left( \bm{\phi} \right)} \right) + \text{dim} \left( {{\text{Null}}\left( \bm{\phi} \right)} \right), 
        \label{R-N}
         \end{equation}
        where $\text{dim} \left( {{\text{Domain}}} \right)$ denotes the input dimension of the linear transformation $\bm{\phi}$. So we can give a linear transformation $\mathcal T:{\text{Image}}\left( {\mathbf{A}} \right) \to {\mathbb{R}^a}$, which is defined as $\mathcal T\left( {\mathbf{v}} \right) = {\mathbf{Cv}},{\mathbf{v}} \in {\text{Image}}\left( {\mathbf{A}} \right)$, i.e., the input to this linear transformation is the image space of $\mathbf{A}$. Thus, according to \eqref{R-N}, 
        \begin{equation}
        \text{dim} \left( {{\text{Domain}}} \right) = \text{dim} \left( {{\text{Image}}\left( {\mathbf{A}} \right)} \right) = {\text{Rank}}\left( {\mathbf{A}} \right).
         \label{proof2_1}
         \end{equation}
         The image space of $\mathcal T$ can be expressed as
        \begin{equation}
       {\text{Image}}\left( \mathcal T \right) = \left\{ {{\mathbf{Cv}}|{\mathbf{v}} \in {\text{Image}}\left( {\mathbf{A}} \right)} \right\} = \left\{ {{\mathbf{CAx}}|\forall {\mathbf{x}} \in {\mathbb{R}^a}} \right\},
         \end{equation}
         which means
         \begin{equation}
       \text{dim} \left( {{\text{Image}}\left( \mathcal T \right)} \right) = {\text{Rank}}\left( {{\mathbf{CA}}} \right).
         \label{proof2_2}
         \end{equation}
         As for the kernel of $\mathcal T$
        \begin{equation}
        \begin{aligned}
         {\text{Null}}\left( \mathcal T \right) &= \left\{ {{\mathbf{v}} \in {\text{Image}}\left( {\mathbf{A}} \right)|{\mathbf{Cv}} = 0} \right\}\\&={\text{Image}}\left( {\mathbf{A}} \right) \cap {\text{Null}}\left( {\mathbf{C}} \right).
         \label{proof2_3}
         \end{aligned}
          \end{equation}
      Above all, according to \eqref{R-N}-\eqref{proof2_3}, the following conclusion can be deduced
      \begin{equation}
      {\text{Rank}}\left( {\mathbf{A}} \right) = {\text{Rank}}\left( {{\mathbf{CA}}} \right) + \text{dim} \left( {{\text{Image}}\left( {\mathbf{A}} \right) \cap {\text{Null}}\left( {\mathbf{C}} \right)} \right),
      \end{equation}
      i.e., 
      \begin{equation}
      {\text{Rank}}\left( {{\mathbf{CA}}} \right) = {\text{Rank}}\left( {\mathbf{A}} \right) - \text{dim} \left( {{\text{Null}}\left( {\mathbf{C}} \right) \cap {\text{Image}}\left( {\mathbf{A}} \right)} \right).
      \end{equation}
      
     \begin{lemma}
     \label{Lemma6}
    $\text{Rank}\left({{\frac{{\partial {\bm{\pi }}}}{{\partial {\mathbf{u}}}}}}\right)=M$ \textit{and} ${\text{Image}}\left( {\frac{{\partial {\bm{\pi }}}}{{\partial {\mathbf{u}}}}} \right) =\\ \left\{ {{\mathbf{v}} \in {\mathbb{R}^{M + \text{1}}}|\sum\limits_{i = \text{1}}^{M + \text{1}} {{v_i}}  = \text{0}} \right\}$.
    \end{lemma}
    
    \textit{Proof.}  According to \eqref{mapping1}, for $\frac{{\partial {\pi_i}}}{{\partial {u_j}}}, i=\text{1},\dots,M+\text{1},j=\text{1},\dots,M+\text{1},$ denote $S={{\sum_{j=\text{1}}^{M+\text{1}}e^{u_j}}}$, based on chain rule, if $i=j$, 
    \begin{equation}
        \begin{aligned}
        \frac{{\partial {\pi_i}}}{{\partial {u_j}}} &=
        \frac{e^{u_i}\left( {S-e^{u_i}} \right)}{S^{2}}=\frac{e^{u_i}}{S}\frac{S-e^{u_i}}{S}=\pi_i\left({1-\pi_i}\right),
        \label{proof33_1}
        \end{aligned}
    \end{equation}
    if $i\ne j$,
    \begin{equation}
        \begin{aligned}
        \frac{{\partial {\pi_i}}}{{\partial {u_j}}} &=
        \frac{0- {e^{u_i}\cdot e^{u_j}} }{S^{2}}=-\frac{e^{u_i}}{S}\frac{e^{u_j}}{S}=-\pi_i\pi_j.
        \label{proof33_2}
        \end{aligned}
    \end{equation}
    Based on \eqref{proof33_1} \eqref{proof33_2}, $\frac{{\partial {\pi_i}}}{{\partial {u_j}}}$ can be given as $\frac{{\partial {\pi_i}}}{{\partial {u_j}}}=\pi_i{\delta_{ij}-\pi_i\pi_j}$, where $\delta_{ij}$ represents the Kronecker delta, which is 1 if $i=j$ and 0 otherwise. From the perspective of $\frac{{\partial {\bm{\pi }}}}{{\partial {\mathbf{u}}}}$, the first part $\pi_i{\delta_{ij}}$ takes a non-zero value only when $i=j$, forming a diagonal matrix with the vector $\bm{\pi}$ as its diagonal $\text{diag}\left({\bm{\pi}}\right)$. The second part $-\pi_i\pi_j$ takes values for all $i,j$ and is, in fact, the outer product of the vectot $\bm{\pi}$ with itself: $-\bm{\pi}\bm{\pi}^\text{T}$. Thus, $\frac{{\partial {\bm{\pi }}}}{{\partial {\mathbf{u}}}}$ can be given as
    \begin{equation}
        \frac{{\partial {\bm{\pi }}}}{{\partial {\mathbf{u}}}}=\text{diag}\left({\bm{\pi}}\right)-\bm{\pi}\bm{\pi}^\text{T},
    \end{equation}
    which can be noticed that $\frac{{\partial {\bm{\pi }}}}{{\partial {\mathbf{u}}}}$ is  a symmetric matrix. Denote ${\bf{v}}=[v_\text{1},\dots,v_{M+\text{1}}]\in\mathbb{R}^{M+\text{1}}, v\in \text{Null}\left({\frac{{\partial {\bm{\pi }}}}{{\partial {\mathbf{u}}}}}\right)$, it holds that 
    \begin{equation}
    \text{diag}\left({\bm{\pi}}\right)\mathbf{v}-\bm{\pi}\left({\bm{\pi}^\text{T}\mathbf{v}}\right)={\bf{0}},
    \end{equation}
    for each component $i$ in the vector $\bm{\pi}$, the above equation can be written as
    \begin{equation}
        \pi_iv_i-\pi_i\left({\bm{\pi}^\text{T}\mathbf{v}}\right)=0.
    \label{rank_pi_u}
    \end{equation}
    For each \textit{i}, it holds that $\pi_i>\text{0}$, then \eqref{rank_pi_u} can be written as
    \begin{equation}
        v_i={\bm{\pi}^\text{T}\mathbf{v}},
        \label{Null}
    \end{equation}
    where ${\bm{\pi}^\text{T}\mathbf{v}}$ is a constant. It means that all components of the vector $\mathbf{v}$ must be equal to each other, i.e., $\text{Null}\left({\frac{{\partial {\bm{\pi }}}}{{\partial {\mathbf{u}}}}}\right)=\text{span}\left({\mathbf{\text{1}}}\right)$ and $\text{dim}\left({(\text{Null}\left({\frac{{\partial {\bm{\pi }}}}{{\partial {\mathbf{u}}}}}\right)}\right)=\text{1}$.
    
    According to \eqref{R-N}, $\text{Rank}\left({{\frac{{\partial {\bm{\pi }}}}{{\partial {\mathbf{u}}}}}}\right)=M$.
    According to spectral theorem, for a symmetric matrix, its image space is the orthogonal complement of its kernel space. Thus, ${\text{Image}}\left( {\frac{{\partial {\mathbf{\pi }}}}{{\partial {\mathbf{u}}}}} \right)$ can be given as $ {\text{Image}}\left( {\frac{{\partial {\mathbf{\pi }}}}{{\partial {\mathbf{u}}}}} \right) = \left\{ {{\mathbf{v}} \in {\mathbb{R}^{M + \text{1}}}|\sum\limits_{i = \text{1}}^{M + \text{1}} {{v_i}}  = \text{0}} \right\}$. 
    
    
    Building upon the aforementioned {\bf{Lemma}} \ref{Lemma5} and {\bf{Lemma}} \ref{Lemma6}, the analysis of matrix $\mathbf{J}_f$ can be completed.
    According to \eqref{mapping3}, for $\frac{{\partial {x_m}}}{{\partial {g_j}}}, m=\text{1},\dots,M, j=\text{1},\dots,M+\text{1},$ it holds that $\frac{{\partial {x_m}}}{{\partial {g_j}}} = \left\{ {\begin{array}{*{20}{c}}
    {\text{1}, j \leqslant m,} \\ 
    {\text{0}, j > m.} \end{array}} \right.$ Thus $\frac{{\partial {\mathbf{p}}}}{{\partial {\mathbf{g}}}}$ can be given as
    \begin{equation}
        \frac{{\partial {\mathbf{p}}}}{{\partial {\mathbf{g}}}} = \left( {\begin{array}{*{20}{c}}
  1&0&{...}&0 \\ 
  1&1&{...}&0 \\ 
   \vdots & \vdots & \vdots & \vdots  \\ 
  1&{...}&1&0 
\end{array}} \right) \in {\mathbb{R}^{M \times \left( {M + 1} \right)}},
\label{proof3_1}
    \end{equation} it is evident that ${\text{Rank}}\left( {\frac{{\partial {\mathbf{p}}}}{{\partial {\mathbf{g}}}}} \right) = M$. According to \eqref{R-N}, it holds that ${\text{dim}}\left( {{\text{Null}}\left( {\frac{{\partial {\mathbf{p}}}}{{\partial {\mathbf{g}}}}} \right)} \right) = \text{1}$ and it can be noticed that 
    \begin{equation}
    {\text{Null}}\left( {\frac{{\partial {\mathbf{p}}}}{{\partial {\mathbf{g}}}}} \right) = \left\{ {{\mathbf{v}} \in {\mathbb{R}^{M + 1}}|{\mathbf{v}} = \left[ {0,0,...,v} \right]}, v\in\mathbb R, v\ne 0 \right\}
    \label{proof3_2}
    \end{equation}
    According to \eqref{mapping2}, it is evident that
    \begin{equation}
        \frac{{\partial {\mathbf{g}}}}{{\partial {\bm{\pi }}}} = {L_{\text{a}}}{\mathbf{I}}.
    \label{gpi}
    \end{equation}

    Thus, based on \eqref{Jf} \eqref{gpi}, ${\text{Rank}}\left( {{{\mathbf{J}}_f}} \right)$ can be given as
    \begin{equation}
    \begin{aligned}
        {\text{Rank}}\left( {{{\mathbf{J}}_f}} \right) &= {\text{Rank}}\left( {\frac{{\partial {\mathbf{p}}}}{{\partial {\mathbf{g}}}} \cdot \frac{{\partial {\bm{\pi }}}}{{\partial {\mathbf{u}}}}} \right),
    \end{aligned}
    \end{equation}
    according to {\bf{Lemma}} \ref{Lemma5}, the right-hand side of the above equation can be represented as ${\text{Rank}}\left( {\frac{{\partial {\mathbf{p}}}}{{\partial {\mathbf{g}}}} \cdot \frac{{\partial {\bm{\pi }}}}{{\partial {\mathbf{u}}}}} \right) = {\text{Rank}}\left( {\frac{{\partial {\bm{\pi }}}}{{\partial {\mathbf{u}}}}} \right) - \text{dim} \left( {{\text{Null}}\left( {\frac{{\partial {\mathbf{p}}}}{{\partial {\mathbf{g}}}}} \right) \cap {\text{Image}}\left( {\frac{{\partial {\bm{\pi }}}}{{\partial {\mathbf{u}}}}} \right)} \right)$. Based on {\bf{Lemma}} \ref{Lemma6} and \eqref{proof3_2}, it is evident that 
    \begin{equation}
        {\text{Rank}}\left( {{{\mathbf{J}}_f}} \right)=M.
    \end{equation} 
    \vspace{-1.5em}
    \section{Proof of Proposition \ref{Lemma2}}
    \label{appendixB}
    The update step in $\mathbf{u}$-space can be given as $\Delta {\mathbf{u}} =  - \alpha  \cdot {\nabla _{\mathbf{u}}}Z$, where $- \alpha$ denotes to the learning rate and $\alpha>\text{0}$. Based on {\bf{Lemma}} \ref{Lemma4}, the update step in $\mathbf{p}$-space can be expressed as
    \begin{equation}
        \Delta {\mathbf{p}} = {{\mathbf{J}}_f} \Delta {\mathbf{u}} = {{\mathbf{J}}_f}  \left( { - \alpha   {\mathbf{J}}_f^{\text{T}} {\nabla _{\mathbf{p}}}Z} \right) =  - \alpha   \left( {{{\mathbf{J}}_f}  {\mathbf{J}}_f^{\text{T}}} \right)  {\nabla _{\mathbf{p}}}Z,
    \end{equation}
    thereby, $\nabla _{\mathbf{p}}^{\text{T}}Z \cdot \Delta {\mathbf{p}} =  - \alpha   \nabla _{\mathbf{p}}^{\text{T}}Z  \left( {{{\mathbf{J}}_f} {\mathbf{J}}_f^{\text{T}}} \right)  {\nabla _{\mathbf{p}}}Z$.
    According to {\bf{Lemma}} \ref{Lemma1}, since $\mathbf{J}_f$ has full row rank, ${{{\mathbf{J}}_f} \cdot {\mathbf{J}}_f^{\text{T}}}$ is a positive definite matrix. By the properties of positive definite matrices, we have $\nabla _{\mathbf{p}}^{\text{T}}Z  \left( {{{\mathbf{J}}_f} {\mathbf{J}}_f^{\text{T}}} \right)  {\nabla _{\mathbf{p}}}Z>\text{0}$, i.e., $\nabla _{\mathbf{p}}^{\text{T}}Z \cdot \Delta {\mathbf{p}}<\text{0}$. Therefore, the update step in $\mathbf{p}$-space aligns with the direction of gradient descent. 
    
    \section{Proof of Proposition \ref{Lemma3}}
    \label{appendixC}
    According to {\bf{Lemma}} \ref{Lemma4} and {\bf{Proposition}} \ref{Lemma1} , ${\nabla _{\mathbf{u}}}Z = {\mathbf{J}}_f^{\text{T}} \cdot {\nabla _{\mathbf{p}}}Z$ and ${\mathbf{J}}_f$ has full row rank. Thus, the columns of ${\mathbf{J}}_f^{\text{T}}$ are linearly independent. Therefore, if ${\nabla _{\mathbf{u}}}Z\left( {{{\mathbf{u}}^ * }} \right) = \text{0}$ is satisfied, the equation holds if and only if ${\nabla _{\mathbf{p}}}Z\left( {{{\mathbf{p}}^ * }} \right) = \text{0}$ is satisfied, in which case $f\left( {{{\mathbf{u}}^ * }} \right) = {{\mathbf{p}}^ * }$ holds. 
    
     \section{Proof of Proposition \ref{convengence_proof}}
       \label{appendixD}
       By Assumption \ref{assumption1} and the standard descent lemma for $L_F$-smooth
functions, we have
\begin{equation}
{\cal L}(\boldsymbol{\beta}_{i+1})
\le
{\cal L}(\boldsymbol{\beta}_{i})
+
\nabla {\cal L}(\boldsymbol{\beta}_{i})^T \mathbf{d}_i
+
\frac{L_F}{2}\left\|\mathbf{d}_i\right\|^2.
\end{equation}
Substituting Assumption \ref{assumption2} yields
\begin{align}
{\cal L}(\boldsymbol{\beta}_{i+1})
&\le
{\cal L}(\boldsymbol{\beta}_{i})
-
c_1\left\|\nabla {\cal L}(\boldsymbol{\beta}_{i})\right\|^2
+
\frac{L_F}{2}c_2^2
\left\|\nabla {\cal L}(\boldsymbol{\beta}_{i})\right\|^2 \\
&=
{\cal L}(\boldsymbol{\beta}_{i})
-
\delta
\left\|\nabla {\cal L}(\boldsymbol{\beta}_{i})\right\|^2,
\end{align}
where
\begin{equation}
\delta \triangleq c_1-\frac{L_F}{2}c_2^2 > 0.
\end{equation}
Summing the above inequality from $i=\text{1}$ to $T$ gives
\begin{equation}
\delta
\sum_{i=1}^{T}
\left\|
\nabla {\cal L}(\boldsymbol{\beta}_i)
\right\|^2
\le
{\cal L}(\boldsymbol{\beta}_1)-{\cal L}(\boldsymbol{\beta}_{T+1}).
\end{equation}
Since ${\cal L}(\boldsymbol{\beta})$ is lower bounded, the right-hand side is
finite. Hence,
\begin{equation}
\sum_{i=1}^{\infty}
\left\|
\nabla {\cal L}(\boldsymbol{\beta}_i)
\right\|^2
< \infty.
\end{equation}
Because each term in the above series is nonnegative, we further obtain
\begin{equation}
\lim_{i\rightarrow\infty}
\left\|
\nabla {\cal L}(\boldsymbol{\beta}_i)
\right\|
=0.
\end{equation}
The first-order stationarity of any accumulation point then follows from
the continuity of $\nabla {\cal L}(\boldsymbol{\beta})$.

	\end{appendices}

\balance
\bibliographystyle{IEEEtran}
\bibliography{reference}

@article{introduction_ISAC1,
  author={Liu, Fan and Cui, Yuanhao and Masouros, Christos and Xu, Jie and Han, Tony Xiao and Eldar, Yonina C. and Buzzi, Stefano},
  journal={IEEE J. Sel. Areas Commun.},
  title={Integrated sensing and communications: Toward dual-functional wireless networks for {6G} and beyond},
  year={2022},
  month={Mar.},
  volume={40},
  number={6},
  pages={1728-1767},
  doi={10.1109/JSAC.2022.3156632}
}

@article{introduction_ISAC2,
  author={González-Prelcic, Nuria and Furkan Keskin, Musa and Kaltiokallio, Ossi and Valkama, Mikko and Dardari, Davide and Shen, Xiao and Shen, Yuan and Bayraktar, Murat and Wymeersch, Henk},
  journal={Proc. IEEE},
  title={The integrated sensing and communication revolution for {6G}: Vision, techniques, and applications},
  year={2024},
  month={May},
  volume={112},
  number={7},
  pages={676-723},
  doi={10.1109/JPROC.2024.3397609}
}

@article{introduction_ISAC3,
  author={Wei, Zhiqing and Qu, Hanyang and Wang, Yuan and Yuan, Xin and Wu, Huici and Du, Ying and Han, Kaifeng and Zhang, Ning and Feng, Zhiyong},
  journal={IEEE Internet Things J.},
  title={Integrated sensing and communication signals toward {5G-A} and {6G}: A survey},
  year={2023},
  month={Jan.},
  volume={10},
  number={13},
  pages={11068-11092},
  doi={10.1109/JIOT.2023.3235618}
}

@article{introduction_ISAC4,
  author={Kaushik, Aryan and Singh, Rohit and Dayarathna, Shalanika and Senanayake, Rajitha and Di Renzo, Marco and Dajer, Miguel and Ji, Hyoungju and Kim, Younsun and Sciancalepore, Vincenzo and Zappone, Alessio and Shin, Wonjae},
  journal={IEEE Commun. Stand. Mag.},
  title={Toward integrated sensing and communications for {6G}: Key enabling technologies, standardization, and challenges},
  year={2024},
  month={May},
  volume={8},
  number={2},
  pages={52-59},
  doi={10.1109/MCOMSTD.0007.2300043}
}

@article{ISAC2,
  author={Su, Nanchi and Liu, Fan and Masouros, Christos},
  journal={IEEE Trans. Wireless Commun.},
  title={Secure radar-communication systems with malicious targets: Integrating radar, communications and jamming functionalities},
  year={2021},
  month={Jan.},
  volume={20},
  number={1},
  pages={83-95},
  doi={10.1109/TWC.2020.3023164}
}

@article{MA1,
  author={Zhu, Lipeng and Ma, Wenyan and Zhang, Rui},
  journal={IEEE Commun. Mag.},
  title={Movable antennas for wireless communication: Opportunities and challenges},
  year={2024},
  month={Jun.},
  volume={62},
  number={6},
  pages={114-120},
  doi={10.1109/MCOM.001.2300212}
}

@article{MA3,
  author={Shao, Xiaodan and Mei, Weidong and You, Changsheng and Wu, Qingqing and Zheng, Beixiong and Wang, Cheng-Xiang and Li, Junling and Zhang, Rui and Schober, Robert and Zhu, Lipeng and Zhuang, Weihua and Shen, Xuemin},
  journal={IEEE Commun. Surveys \& Tuts.},
  title={A tutorial on six-dimensional movable antenna for {6G} networks: Synergizing positionable and rotatable antennas},
  year={2025},
  month={Aug.},
  volume={28},
  pages={3666-3709},
  doi={10.1109/COMST.2025.3602939}
}

@article{MAPosition,
  author={Mei, Weidong and Wei, Xin and Ning, Boyu and Chen, Zhi and Zhang, Rui},
  journal={IEEE Wireless Commun. Lett.},
  title={Movable-antenna position optimization: A graph-based approach},
  year={2024},
  month={Jul.},
  volume={13},
  number={7},
  pages={1853-1857},
  doi={10.1109/LWC.2024.3393135}
}

@article{MA_jointopt1,
  author={Wei, Xin and Mei, Weidong and Wang, Dong and Ning, Boyu and Chen, Zhi},
  journal={IEEE Wireless Commun. Lett.},
  title={Joint beamforming and antenna position optimization for movable antenna-assisted spectrum sharing},
  year={2024},
  month={Sep.},
  volume={13},
  number={9},
  pages={2502-2506},
  doi={10.1109/LWC.2024.3421636}
}

@inproceedings{MA_jointopt2,
  author={Chen, Xintai and Feng, Biqian and Wu, Yongpeng and Kwan Ng, Derrick Wing and Schober, Robert},
  booktitle={Proc. IEEE Global Commun. Conf. (GLOBECOM)},
  title={Joint beamforming and antenna movement design for moveable antenna systems based on statistical {CSI}},
  year={2023},
  month={Feb.},
  pages={4387-4392},
  doi={10.1109/GLOBECOM54140.2023.10437006}
}

@inproceedings{MA_jointopt3,
  author={Wu, Yifei and Xu, Dongfang and Ng, Derrick Wing Kwan and Gerstacker, Wolfgang and Schober, Robert},
  booktitle={Proc. IEEE Global Commun. Conf. (GLOBECOM)},
  title={Movable antenna-enhanced multiuser communication: Jointly optimal discrete antenna positioning and beamforming},
  year={2023},
  month={Feb.},
  pages={7508-7513},
  doi={10.1109/GLOBECOM54140.2023.10437926}
}

@article{MAISAC1,
  author={Li, Zhendong and Ba, Jianle and Su, Zhou and Huang, Jinyuan and Peng, Haixia and Chen, Wen and Du, Linkang and Luan, Tom H.},
  journal={IEEE Wireless Commun.},
  title={Movable antennas enabled {ISAC} systems: Fundamentals, opportunities, and future directions},
  year={2025},
  month={Nov.},
  volume={25},
  pages={7220-7234},
  doi={10.1109/TWC.2025.3630154}
}

@article{MAISAC2,
  author={Qin, Haoran and Chen, Wen and Wu, Qingqing and Zhang, Ziheng and Li, Zhendong and Cheng, Nan},
  journal={IEEE Wireless Commun. Lett.},
  title={{Cram{\'e}r-Rao} bound minimization for movable antenna-assisted multiuser integrated sensing and communications},
  year={2024},
  month={Dec.},
  volume={13},
  number={12},
  pages={3404-3408},
  doi={10.1109/LWC.2024.3468709}
}

@article{MAISAC3,
  author={Chen, Lebin and Zhao, Ming-Min and Zhao, Min-Jian and Zhang, Rui},
  journal={IEEE Trans. Signal Process.},
  title={Antenna position and beamforming optimization for movable antenna enabled {ISAC}: Optimal solutions and efficient algorithms},
  year={2025},
  month={Jul.},
  volume={73},
  pages={3812-3828},
  doi={10.1109/TSP.2025.3590379}
}

@article{MAISAC5,
  author={Li, Zhendong and Ba, Jianle and Su, Zhou and Peng, Haixia and Wang, Yuntao and Chen, Wen and Wu, Qingqing},
  journal={IEEE Trans. Wireless Commun.},
  title={Joint discrete antenna positioning and beamforming optimization in movable antenna enabled full-duplex {ISAC} networks},
  year={2025},
  month={Nov.},
  volume={25},
  pages={7220-7234},
  doi={10.1109/TWC.2025.3630154}
}

@article{MAlow1,
  author={Xiu, Yue and Yang, Songjie and Lyu, Wanting and Lep Yeoh, Phee and Li, Yonghui and Ai, Yi},
  journal={IEEE Wireless Commun. Lett.},
  title={Movable antenna enabled {ISAC} beamforming design for low-altitude airborne vehicles},
  year={2025},
  month={May},
  volume={14},
  number={5},
  pages={1311-1315},
  doi={10.1109/LWC.2025.3538587}
}

@inproceedings{MAlow2,
  author={Kuang, Ziming and Liu, Wenchao and Wang, Chunjie and Jin, Zhenzhen and Ren, Jinke and Zhang, Xuhui and Shen, Yanyan},
  booktitle={Proc. IEEE/CIC Int. Conf. Commun. China (ICCC Workshops)},
  title={Movable-antenna array empowered {ISAC} systems for low-altitude economy},
  year={2024},
  month={Oct.},
  pages={776-781},
  doi={10.1109/ICCCWorkshops62562.2024.10693833}
}

@article{MA2,
  author={Zhu, Lipeng and Ma, Wenyan and Zhang, Rui},
  journal={IEEE Trans. Wireless Commun.},
  title={Modeling and performance analysis for movable antenna enabled wireless communications},
  year={2024},
  month={Jun.},
  volume={23},
  number={6},
  pages={6234-6250},
  doi={10.1109/TWC.2023.3330887}
}

@article{MASecure,
  author={Hu, Guojie and Wu, Qingqing and Xu, Kui and Si, Jiangbo and Al-Dhahir, Naofal},
  journal={IEEE Signal Process. Lett.},
  title={Secure wireless communication via movable-antenna array},
  year={2024},
  month={Jan.},
  volume={31},
  pages={516-520},
  doi={10.1109/LSP.2024.3359894}
}

@article{MAdif1,
  author={Ma, Wenyan and Zhu, Lipeng and Zhang, Rui},
  journal={IEEE Trans. Wireless Commun.},
  title={{MIMO} capacity characterization for movable antenna systems},
  year={2024},
  month={Apr.},
  volume={23},
  number={4},
  pages={3392-3407},
  doi={10.1109/TWC.2023.3307696}
}

@article{MADRL,
  author={Wang, Chao and Li, Guo and Zhang, Haibin and Wong, Kai-Kit and Li, Zan and Ng, Derrick Wing Kwan and Chae, Chan-Byoung},
  journal={IEEE Trans. Wireless Commun.},
  title={Fluid antenna system liberating multiuser {MIMO} for {ISAC} via deep reinforcement learning},
  year={2024},
  month={Sep.},
  volume={23},
  number={9},
  pages={10879-10894},
  doi={10.1109/TWC.2024.3376800}
}

@article{MADL,
  author={Qi, Qiao and Chen, Xiaoming and Zhong, Caijun and Yuen, Chau and Zhang, Zhaoyang},
  journal={IEEE Trans. Wireless Commun.},
  title={Deep learning-based design of uplink integrated sensing and communication},
  year={2024},
  month={Sep.},
  volume={23},
  number={9},
  pages={10639-10652},
  doi={10.1109/TWC.2024.3373797}
}

@article{WMMSE,
  author={Shi, Qingjiang and Razaviyayn, Meisam and Luo, Zhi-Quan and He, Chen},
  journal={IEEE Trans. Signal Process.},
  title={An iteratively weighted {MMSE} approach to distributed sum-utility maximization for a {MIMO} interfering broadcast channel},
  year={2011},
  month={Sep.},
  volume={59},
  number={9},
  pages={4331-4340},
  doi={10.1109/TSP.2011.2147784}
}

@inproceedings{Meta1,
  author={Xia, Jingyuan and Gunduz, Deniz},
  booktitle={Proc. IEEE Int. Symp. Inf. Theory (ISIT)},
  title={Meta-learning based beamforming design for {MISO} downlink},
  year={2021},
  pages={2954-2959},
  doi={10.1109/ISIT45174.2021.9518251}
}

@article{Meta2,
  author={Xia, Jing-Yuan and Li, Shengxi and Huang, Jun-Jie and Yang, Zhixiong and Jaimoukha, Imad M. and Gündüz, Deniz},
  journal={IEEE Trans. Neural Netw. Learning Syst.},
  title={Metalearning-based alternating minimization algorithm for nonconvex optimization},
  year={2023},
  month={Sep.},
  volume={34},
  number={9},
  pages={5366-5380},
  doi={10.1109/TNNLS.2022.3165627}
}

@article{Meta4,
  author={Zhu, Fenghao and Wang, Xinquan and Huang, Chongwen and Yang, Zhaohui and Chen, Xiaoming and Al Hammadi, Ahmed and Zhang, Zhaoyang and Yuen, Chau and Debbah, Mérouane},
  journal={IEEE Trans. Wireless Commun.},
  title={Robust beamforming for {RIS}-aided communications: Gradient-based manifold meta learning},
  year={2024},
  month={Nov.},
  volume={23},
  number={11},
  pages={15945-15956},
  doi={10.1109/TWC.2024.3435023}
}

@article{Rd,
  author={Liu, Fan and Masouros, Christos and Li, Ang and Sun, Huafei and Hanzo, Lajos},
  journal={IEEE Trans. Wireless Commun.},
  title={{MU-MIMO} communications with {MIMO} radar: From co-existence to joint transmission},
  year={2018},
  month={Apr.},
  volume={17},
  number={4},
  pages={2755-2770},
  doi={10.1109/TWC.2018.2803045}
}

@article{SECUREISAC1,
  author={G{\"u}nl{\"u}, Onur and Bloch, Matthieu R. and Schaefer, Rafael F. and Yener, Aylin},
  journal={IEEE J. Sel. Areas Inf. Theory},
  title={Secure integrated sensing and communication},
  year={2023},
  month={May},
  volume={4},
  pages={40-53},
  doi={10.1109/JSAIT.2023.3275048}
}

@inproceedings{SECUREISAC4,
  author={Liu, Yutong and Zhu, Zengbao and Cui, Qimei and Duo, Hao},
  booktitle={Proc. IEEE Int. Conf. Commun. Workshops (ICC Workshops)},
  title={Artificial-noise-aided secure transmit beamforming for {MU-MISO} integrated sensing and communication systems},
  year={2024},
  month={Aug.},
  pages={1219-1224},
  doi={10.1109/ICCWorkshops59551.2024.10615699}
}

@article{SECUREISAC8,
  author={Dan, Qian and Lei, Hongjiang and Park, Ki-Hong and Pan, Gaofeng},
  journal={IEEE Trans. Cogn. Commun. Netw.},
  title={Beamforming for secure {RSMA}-aided {ISAC} systems},
  year={2025},
  month={July},
  volume={11},
  number={5},
  pages={2970-2983},
  doi={10.1109/TCCN.2025.3587100}
}

@article{SECUREISAC9,
  author={Zhang, Lei and Wang, Yinghui and Chen, Hang and Cao, Yang},
  journal={IEEE Internet Things J.},
  title={Physical-layer security of the {NOMA}-assisted {ISAC} systems under near-field scenario},
  year={2025},
  month={Feb.},
  volume={12},
  number={12},
  pages={18546-18553},
  doi={10.1109/JIOT.2025.3543366}
}

@inproceedings{SECUREISAC10,
  author={Naim Shaikh, Mohd Hamza and Celik, Abdulkadir and Eltawil, Ahmed M. and Nauryzbayev, Galymzhan},
  booktitle={Proc. Asilomar Conf. Signals Syst. Comput.},
  title={Sense and jam: {ISAC}-aided robust physical layer security},
  year={2024},
  month={Apr.},
  pages={653-657},
  doi={10.1109/IEEECONF60004.2024.10942975}
}

@article{SECUREISAC11,
  author={Su, Nanchi and Liu, Fan and Masouros, Christos},
  journal={IEEE Trans. Wireless Commun.},
  title={Sensing-assisted eavesdropper estimation: An {ISAC} breakthrough in physical layer security},
  year={2024},
  month={Aug.},
  volume={23},
  number={4},
  pages={3162-3174},
  doi={10.1109/TWC.2023.3306029}
}

@article{SECUREISAC12,
  author={Chen, Chaodi and Yao, Junteng and Jin, Ming and Guo, Qinghua},
  journal={IEEE Wireless Commun. Lett.},
  title={Beamforming and computing capacity allocation for {ISAC}-assisted secure mobile edge computing},
  year={2024},
  month={Sep.},
  volume={13},
  number={12},
  pages={3360-3364},
  doi={10.1109/LWC.2024.3463752}
}

@article{SECUREISAC14,
  author={Guo, Liang and Jia, Jie and Chen, Jian and Yang, Shuhui and Wang, Xingwei},
  journal={IEEE Trans. Inf. Forensics Security},
  title={Secure beamforming and radar association in {CoMP}-{NOMA} empowered integrated sensing and communication systems},
  year={2024},
  month={Oct.},
  volume={19},
  pages={10246-10257},
  doi={10.1109/TIFS.2024.3484950}
}

@ARTICLE{MAhardware1,
  author={Dong, Zhenjun and Zhou, Zhiwen and Xiao, Zhiqiang and Zhang, Chaoyue and Li, Xinrui and Min, Hongqi and Zeng, Yong and Jin, Shi and Zhang, Rui},
  journal={IEEE Trans. Wireless Commun.}, 
  title={Movable Antenna for Wireless Communications: Prototyping and Experimental Results}, 
  year={2025},
  month={Nov.},
  volume={25},
  pages={6586-6599},
  doi={10.1109/TWC.2025.3625559}}

@article{MAhardware2,
  author={Ning, Boyu and Yang, Songjie and Wu, Yafei and Wang, Peilan and Mei, Weidong and Yuen, Chau and Bj{\"o}rnson, Emil},
  journal={IEEE Wireless Commun.}, 
  title={Movable antenna-enhanced wireless communications: General architectures and implementation methods}, 
  year={2025},
  month={May},
  volume={32},
  number={5},
  pages={108-116},
  doi={10.1109/MWC.013.2400238}
}

@article{Tang2025Deep,
  author={Tang, Xiao and Zhao, Kexin and Shen, Chao and Du, Qinghe and Wang, Yichen and Niyato, Dusit and Han, Zhu},
  journal={IEEE Trans. Mobile Comput.}, 
  title={Deep graph reinforcement learning for {UAV}-enabled multi-user secure communications}, 
  year={2025},
  month={Apr.},
  volume={24},
  number={9},
  pages={8780-8793},
  doi={10.1109/TMC.2025.3558790}
}

@article{FAS1,
  author={Yao, Junteng and Xin, Liangxiao and Wu, Tuo and Jin, Ming and Wong, Kai-Kit and Yuen, Chau and Shin, Hyundong},
  journal={IEEE Internet Things J.}, 
  title={{FAS} for secure and covert communications}, 
  year={2025},
  month={Jun.},
  volume={12},
  number={11},
  pages={18414-18418},
  doi={10.1109/JIOT.2025.3539737}
}

@article{MAISAC_new1,
  author={Lyu, Wanting and Dong, Xiangyu and Yang, Ran and Wang, Kaihe and Zhang, Zhongpei and Assi, Chadi and Yuen, Chau},
  journal={IEEE Trans. Wireless Commun.}, 
  title={{NOMA}-empowered integrated sensing and communication with movable antennas}, 
  year={2025},
  month={Dec.},
  volume={25},
  pages={8401-8416},
  doi={10.1109/TWC.2025.3637741}
}

@article{MAISAC_new2,
  author={Jiang, Chengjun and Zhang, Chensi and Huang, Chongwen and Ge, Jianhua and Niyato, Dusit and Yuen, Chau},
  journal={IEEE Trans. Wireless Commun.}, 
  title={Movable antenna-assisted integrated sensing and communication systems}, 
  year={2025},
  month={Aug.},
  volume={24},
  number={8},
  pages={6397-6412},
  doi={10.1109/TWC.2025.3552843}
}

@article{FAS2,
  author={Yang, Shunxing and Yao, Junteng and Tang, Jie and Wu, Tuo and Elkashlan, Maged and Yuen, Chau and Debbah, Mérouane and Shin, Hyundong and Valenti, Matthew},
  journal={IEEE Internet Things J.}, 
  title={Toward intelligent antenna positioning: Leveraging {DRL} for {FAS}-aided {ISAC} systems}, 
  year={2025},
  month={Aug.},
  volume={12},
  number={16},
  pages={34615-34618},
  doi={10.1109/JIOT.2025.3576235}
}

@article{MVPSO,
  author={Ding, Jingze and Zhou, Zijian and Jiao, Bingli},
  journal={IEEE Trans. Wireless Commun.}, 
  title={Movable antenna-aided secure full-duplex multi-user communications}, 
  year={2025},
  month={Mar.},
  volume={24},
  number={3},
  pages={2389-2403},
  doi={10.1109/TWC.2024.3520806}
}

\end{document}